\documentclass[aps,showpacs,preprintnumbers,amsmath,amssymb,
eqsecnum, twocolumn, tightenlines]{revtex4}
\usepackage{graphicx}
\usepackage[usenames,dvipsnames]{color}

\DeclareMathOperator\arctanh{arctanh}

\newcommand{\be}{\begin{eqnarray}}
\newcommand{\ee}{\end{eqnarray}}

 \newcommand{\gsim}{\mathrel{\hbox{\rlap{\lower.55ex \hbox {$\sim$}}
                   \kern-.3em \raise.4ex \hbox{$>$}}}}
\newcommand{\lsim}{\mathrel{\hbox{\rlap{\lower.55ex \hbox {$\sim$}}
                   \kern-.3em \raise.4ex \hbox{$<$}}}}

\newcommand{\ba}{\begin{eqnarray}}
\newcommand{\ea}{\end{eqnarray}}
\newcommand{\nn}{\nonumber}
\def\bea{\be}
\def\eea{\ee}

\def\roughly#1{\mathrel{\raise.3ex\hbox{$#1$\kern-.75em%
\lower1ex\hbox{$\sim$}}}}
\def\lsim{\roughly<}
\def\gsim{\roughly>}

\def\({\left(}
\def\){\right)}
\def\[{\left[}
\def\]{\right]}

\def\lsim{\mathrel{\rlap{\lower3pt\hbox{\hskip1pt$\sim$}}
     \raise1pt\hbox{$<$}}} %less than or approx. symbol
\def\gsim{\mathrel{\rlap{\lower3pt\hbox{\hskip1pt$\sim$}}
     \raise1pt\hbox{$>$}}} %greater than or approx. symbol

\def\lab{\label}
\def\le{\left}
\def\ri{\right}
\def\bea{\begin{eqnarray}}
\def\eea{\end{eqnarray}}

\def\II{\relax{\rm I\kern-.18em I}}

\def\l{\lambda}
\def\m{\mu}
\def\n{\nu}
\def\r{\rho}
\def\s{\sigma}

\def\xp{x^{+}}

\def\f{\varphi}

\def\lab{\label}

\def\le{\left}
\def\ri{\right}

\def\6{\partial}

\def\nn{\nonumber}

\def\f{\phi}

\def\half{\frac12}
\def\xp{x_\perp}

\begin{document}

%\twocolumn[\hsize\textwidth\columnwidth\hsize\csname @twocolumnfalse\endcsname

\title{Magnetohydrodynamics, charged currents and directed flow in heavy ion collisions}

\author { Umut G\"ursoy$^1$, Dmitri Kharzeev$^{2,3}$ and Krishna Rajagopal$^4$ }
\affiliation { $^1$ Institute for Theoretical Physics, Utrecht University\\ Leuvenlaan 4, 3584 CE Utrecht, The Netherlands
}
\affiliation { $^2$ Department of Physics and Astronomy, Stony Brook University, New York 11794, USA}
\affiliation { $^3$ Department of Physics, Brookhaven National Laboratory, Upton, New York 11973, USA}
\affiliation { $^4$ Center for Theoretical Physics, Massachusetts Institute of Technology, Cambridge, MA 02139.}
\date{\today}

\begin{abstract}
The hot QCD matter produced in any heavy ion collision with a nonzero impact
parameter is produced within a strong magnetic field.
We study the imprint that these fields leave on the azimuthal distributions and 
correlations of the produced charged hadrons. 
The magnetic field is time-dependent and the medium is expanding, which leads to the induction of 
charged currents due to the combination of Faraday and Hall effects. We find that these currents 
result in a charge-dependent
directed flow $v_1$ that is odd in rapidity and odd under charge exchange. 
It can be detected by measuring
correlations between the directed flow of charged hadrons at different rapidities, 
$\langle v_1^\pm (y_1) v_1^\pm (y_2) \rangle$. 
\end{abstract}

\preprint{MIT-CTP-4526}

%\vskip0.1cm
%We study the imprint of magnetic field, B that is present in the non-central heavy ion collisions on the hadron spectra. Magnetic field is time-dependent and the medium is expanding. Therefore one expects production of charged currents due to both Faraday and Hall effects. We argue that these charged currents in the QGP have a direct impact on the directed flow parameter $v_1$. We model the expanding plasma by conformal hydrodynamics solution of Gubser \cite{Gubser} and superpose magnetic field on this solution in a perturbative manner. Directed flow is calculated by a Cooper-Frye analysis and we indeed observe non-trivial effects of B on $v_1$.         
%\end{abstract}
\maketitle

%\documentstyle[11pt,amssymb,epsfig,amsmath]{article}\def\baselinestretch{1.4}
%%\usepackage{graphicx}
%%\documentstyle[11pt,psfig]{article}\def\baselinestretch{1.2}

%\parindent 30pt\textheight 9in\topmargin -.7in\textwidth 6in
%\oddsidemargin .25in\evensidemargin 0in
%\begin{document}
%\newcommand{\be}{\begin{equation}}
%\newcommand{\ee}{\end{equation}}
%\newcommand{\ba}{\begin{eqnarray}}
%\newcommand{\ea}{\end{eqnarray}}
%\newcommand{\no}{\nonumber \\}
%\newcommand{\gsim}{\mathrel{\hbox{\rlap{\lower.55ex \hbox {$\sim$}}
%                   \kern-.3em \raise.4ex \hbox{$>$}}}}
%\newcommand{\lsim}{\mathrel{\hbox{\rlap{\lower.55ex \hbox {$\sim$}}
%                   \kern-.3em \raise.4ex \hbox{$<$}}}}

%
%\def\be{\begin{eqnarray}}
%\def\ee{\end{eqnarray}}

%\renewcommand{\thefootnote}{\arabic{footnote}}
%\setcounter{footnote}{0}

%\vskip 0.4cm \hfill { }
% \hfill {\today} \vskip 1cm

%\begin{center}
%{\LARGE\bf Notes on scalar/vector perturbations on KO background
%   }
%\date{\today}

%\vskip 1cm {\large %Shu
%%Lin\footnote{E-mail:slin@grad.physics.sunysb.edu},
%%and Edward
%%Shuryak\footnote{E-mail:shuryak@tonic.physics.sunysb.edu}
% }

%%{\large and}

%\end{center}

%\vskip 0.5cm

%\begin{center}

%

%

%
%\end{center}

%\vskip 0.5cm

%\begin{abstract}
%\end{abstract}

%\newpage

%\renewcommand{\thefootnote}{\#\arabic{footnote}}
%\setcounter{footnote}{0}

\section{Introduction}
\lab{intro}

Strong magnetic fields $\vec B$ are produced 
in all non-central 
heavy ion collisions (i.e.~those with nonzero impact parameter $b$) by the charged ``spectators'' (i.e. the nucleons from the incident nuclei that ``miss'',
flying past each other rather than colliding). 
Indeed,  estimates obtained via 
application of the Biot-Savart law to heavy ion collisions with
$b=4$~fm yield $e|\vec B|/m_\pi^2 \approx$ 1-3 about 0.1-0.2 fm$/c$
after a RHIC collision with $\sqrt{s}=200$~AGeV 
and $e|\vec B|/
m_\pi^2 \approx $ 10-15 at some even earlier time after 
an LHC collision with $\sqrt{s}=2.76$~ATeV~\cite{Kharzeev:2007jp,Skokov:2009qp,Tuchin1,Voronyuk:2011jd,Deng:2012pc,Tuchin2,McLerran:2013hla}. 
In recent years there has been much interest in consequences of these 
enormous magnetic fields present early in the collision that are observable
in the final state hadrons produced by the collision.  In particular,
the interplay of magnetic field and quantum anomalies has been predicted to lead to a number of interesting phenomena, including the chiral magnetic effect \cite{Kharzeev:2007jp,Fukushima:2008xe}, 
a quadrupole deformation of the electric charge distribution induced by a chiral magnetic 
wave~\cite{Kharzeev:2010gd,Burnier:2011bf}, and the enhanced anisotropic production of soft 
photons through  ``magneto-sonoluminescence" -- the conversion of phonons into photons in an external 
magnetic field \cite{Basar:2012bp}. 
While several of the predicted effects have been observed in 
heavy ion collision data~\cite{Abelev:2009ac,Abelev:2009ad,Abelev:2012pa,Wang:2012qs,Drees:2013wza,Adamczyk:2013hsi,Adamczyk:2013kcb}, 
it is often hard to distinguish them unambiguously from a combination of mundane 
phenomena possibly present in the anisotropic expansion of quark-gluon matter, see 
e.g. Refs.~\cite{Schlichting:2010qia,Bzdak:2012ia,Bzdak:2013yla}. 
This makes it imperative to establish that the presence of an early-time magnetic
field can have observable consequences on the motion of
the final-state charged particles seen in detectors, 
making it possible to use data 
to calibrate the strength of the magnetic field.

%Because of this, it is necessary to establish the presence of magnetic field 
%and to calibrate its strength using its effect on the production of charged particles. 

\begin{figure}[t!]
 \begin{center}
\includegraphics[scale=0.38]{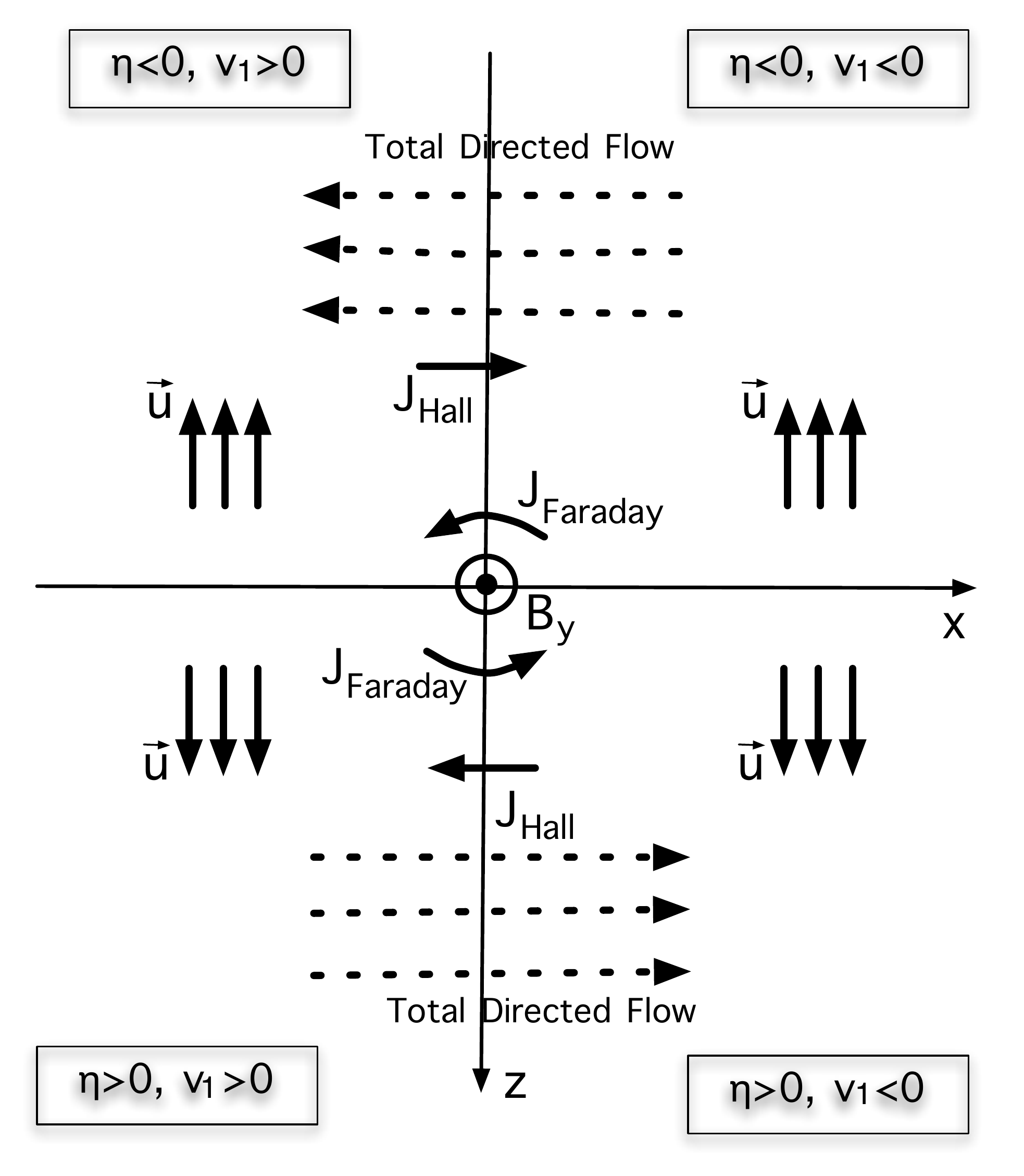}
 \end{center}
 \caption[]{Schematic illustration of how the magnetic field $\vec B$ in a heavy ion collision results
in a directed flow, $v_1$, of electric charge.  The collision occurs in the $z$-direction, meaning that
the longitudinal expansion velocity $\vec u$ of the conducting QGP that is produced in
the collision points in the $+z$ ($-z$)
direction at positive (negative) $z$.  We take the impact parameter vector to point in the $x$ direction,
choosing the nucleus moving toward positive (negative) $z$ to be located at negative (positive) $x$,
which is to say taking
the magnetic field $\vec B$ to point in the $+y$ direction.
The direction of the electric currents %fields 
due to the Faraday and Hall effects is shown,
as is the direction of the directed flow of positive charge (dashed) in the case where
the Faraday effect is on balance stronger than the Hall effect. 
In some regions
of spacetime, the electric current %field 
due to the Hall effect is greater than that due to
the Faraday effect; in other regions, the Faraday-induced current %field 
is stronger.
The computation of the directed flow of charged particles is a suitably weighted
integral over spacetime, meaning that the final result for the directed flow 
arises from a partial cancellation between the opposing Faraday and Hall effects.
In some settings (i.e. for some hadron species, with momenta in some ranges)
the total directed flow for positively charged particles points as shown.  In other
settings, it points  in the opposite direction.
}
\label{fig1}
\end{figure}

In this paper we analyze what are surely the simplest and most direct effects
of magnetic fields in heavy ion collisions, and quite likely also their largest effects,
namely the induction of electric currents carried by the charged quarks and antiquarks
in the quark-gluon plasma (QGP) and, later, by the charged hadrons.
The source of these charged currents is twofold. 
Firstly, the magnitude of $\vec B$ varies in time, 
decreasing as the charged spectators fly away along the beam direction, receding
from the QGP produced in the collision.  The changing $\vec B$ results in an electric field
due to Faraday's law, and this in turn produces an electric current in the conducting
medium.  Secondly, because the conducting medium, i.e.~the QGP, has a significant
initial longitudinal expansion velocity $\vec u$ parallel to the beam direction and
therefore perpendicular to $\vec B$, the Lorentz force results in 
%an electric field and a consequent 
an electric current perpendicular to both the velocity and $\vec B$,
akin to the classical Hall effect. (We shall refer to this current
as a Hall current throughout, even though this nomenclature
may not be quite right since our system has no edge at which
charges can build up.)
%The second source stems from the expansion of the QGP -- this is the Hall effect: when magnetic field is applied to a conducting medium the current acquires a component perpendicular to the electric field. 
%As a consequence of Lorentz invariance, a similar effect is produced in an expanding plasma in the lab frame. 
Fig.~\ref{fig1} serves to orient the reader as to the directions of $\vec B$ and 
$\vec u$, and the electric currents induced by the Faraday and Hall effects.
The net electric current
%(i.e.~
is
the sum of that due to Faraday and that due to Hall.
%)
%will drive an electric current in the conducting plasma.  
If the Faraday effect is stronger
than the Hall effect, that current will result in directed flow of positively charged particles
in the directions shown in Fig.~\ref{fig1} and directed flow of negatively charged particles in
the opposite direction.  Our goal in this paper is to make an estimate of the
order of magnitude of the resulting charge-dependent $v_1$ in the final state
pions and (anti)protons.    We will make many simplifying assumptions, since our goal 
is only to show which $v_1$-correlations can be used to look for effects
of the initial magnetic field and to give experimentalists an order-of-magnitude
sense of how large these correlations  may reasonably be expected to be.

%It is reasonable to expect that presence of charged currents would result in a non-trivial directed flow coefficient $v_1$ in the observed charged hadron  spectrum such as the charged pions and protons. 

%%%%%%%%%%%%%%%%%%%%%%%%%%%%%%%%%%%%%%%%%%%%%%%
%%%%%%%%%%%%%%%%%%%%%%%%%%%%%%%%%%%%%%%%%%%%%%%%%%%%%%%%%%%%%%%

The biggest simplifying assumption that we shall make is to treat the electrical conductivity of the 
QGP  $\sigma$ as if it were a constant.  We make this assumption only because it will permit us
to do a mostly analytic calculation.
In reality, $\sigma$ is certainly temperature dependent:
just on dimensional grounds it is expected to be proportional
to the temperature of the plasma.
This means that $\sigma$ should certainly be a function of 
space and time as the  plasma expands and flows hydrodynamically, 
with $\sigma$ decreasing as the plasma cools.  
Furthermore, during the early pre-equilibrium epoch $\sigma$ should rapidly increase
from zero to its equilibrium value.
Taking all this into consideration would require a full, numerical, magnetohydrodynamic
analysis, which we leave to the future.  We shall treat $\sigma$ as a constant, unchanging
until freezeout.
We select a reasonable order-of-magnitude value of the conductivity $\sigma$ based upon
recent lattice calculations~\cite{Ding:2010ga,Francis:2011bt,Brandt:2012jc,Amato:2013naa,Kaczmarek:2013dya}.
It is conventional in these calculations to quote results for $C_{\rm em}^{-1} \sigma / T$, where
$C_{\rm em}\equiv (\frac{4}{9} + \frac{1}{9} + \frac{1}{9})e^2 = 0.061$ in 3-flavor QCD.  
The quantity $C_{\rm em}^{-1} \sigma / T$ is weakly temperature dependent between about $1.2 T_c$
and $2 T_c$, with $T_c \sim 170$~MeV 
the temperature of the crossover from a hadron gas to quark-gluon plasma.  At $T=1.5 T_c \sim 255$~MeV,
$C_{\rm em}^{-1} \sigma / T$ lies between 
0.2 and 0.4~\cite{Ding:2010ga,Francis:2011bt,Brandt:2012jc,Amato:2013naa,Kaczmarek:2013dya}.
We shall set $\sigma=0.023$~fm$^{-1}$ throughout this paper. This corresponds to $C_{\rm em}^{-1} \sigma / T = 0.3$
at $T=255$~MeV.

To do an analytic calculation we need an analytic solution for the hydrodynamic
expansion of the conducting fluid in the absence of any electric currents.
We shall use the analytic solution to relativistic viscous hydrodynamics for a conformal fluid
with the shear viscosity to entropy density ratio given by $\eta/s=1/(4\pi)$ 
found by Gubser in 2010~\cite{Gubser}. 
The solution describes a finite size plasma produced in a {\em central} collision that is obtained from 
conformal hydrodynamics by demanding boost invariance along the beam (i.e.~$z$) direction, rotational invariance around $z$, and two special conformal invariances perpendicular to $z$. This leads to a fluid flow 
that preserves a $SO(1,1)\times SO(3)\times Z_2$ subgroup of the full $4$-dimensional conformal group, 
with the $Z_2$ coming from invariance under $z \leftrightarrow -z$.
Gubser obtains analytic expressions for the four-velocity $u^\mu$ from which one can construct the 
local temperature and energy density of the conformal fluid.  As we demonstrate below,
we can choose parameters such that Gubser's solution yields
a reasonable facsimile of the  pion and proton transverse momentum spectra observed
in RHIC and LHC collisions with $20-30\%$ centrality, corresponding to collisions with
a mean impact parameter between $7$ and $8$~fm, see e.g. \cite{Kharzeev:2000ph,Kharzeev:2004if}.  Gubser's hydrodynamic solution
is rotationally invariant around the $z$-direction and so in reality cannot be directly applicable
to collisions with nonzero impact parameter.  A future numerical
analysis should be based instead upon a numerical solution to  (3+1)-dimensional relativistic
hydrodynamics for non-central heavy ion collisions.

%Even though this solution requires a rotational invariance along the beam axis and hence is not directly applicable to non-central collisions, we argue below that it can be used as a background also in the presence of a non-central collision 

We shall assume throughout that the effects of the magnetic field are small in the sense that
the velocity of charged particles that results (via Hall and Faraday) 
from the presence of $\vec B$, call it $\vec v$,
is much smaller than the velocity of the expanding plasma $\vec u$.
That is, we require $|\vec{v}| \ll  |\vec{u}| $. 
We shall see that this is a good assumption.
Upon making this assumption, and given that our goal is only an order-of-magnitude
estimate of the magnitude of the charge-dependent directed flow,  all we really need from
hydrodynamics is a flow field $\vec u$ that is 
reasonable in transverse extent and in magnitude, 
and a temperature field $T$ that can be used
to define a reasonable freezeout surface in spacetime
at which the hydrodynamic fluid cools below some
specified freezeout temperature and is replaced by hadrons,
following the Cooper-Frye procedure~\cite{Cooper:1974mv}.
In particular, we shall only be interested
in the small charge-dependent azimuthal anisotropy $v_1$
due to the velocity $\vec v$ of charged particles and shall not be interested
at all in the larger, but charge-independent, azimuthal anisotropies
in the hydrodynamic expansion that are induced by the initial azimuthal
anisotropy in collisions with nonzero impact parameter.
For all our purposes, therefore, Gubser's azimuthally symmetric solution
suffices.

In order to obtain the velocity $\vec v$ 
associated with the charged currents due to the electromagnetic field, in Section 2 we first calculate the magnetic and electric fields
themselves, $\vec{B}$ and $\vec{E}$, by solving Maxwell's equations 
in the center-of-mass frame (the frame illustrated in Fig.~\ref{fig1}).  
From $\vec{E}$ and the electrical conductivity $\sigma$ it would be
straightforward to obtain the electric current density $\vec J = \sigma \vec E$.
However, for our purposes what we need is not $\vec J$ itself. 
The electric current $\vec J$ will be associated
with positively charged fluid moving with mean velocity $\vec v$
and negatively charged fluid moving with mean velocity $-\vec v$, 
and what we need to determine is the magnitude and direction of $\vec v$.

In order to determine $\vec v$ at some point in spacetime, we first
boost to the local fluid rest frame at that point in spacetime, namely the
(primed) frame in which $\vec{u'}=0$ at that point.
%by a Lorentz transformation $\Lambda(\vec{u})$. 
 In the primed
 frame all components of the electromagnetic field $\vec{E'}$ and $\vec{B'}$ are non-vanishing. 
 We then
 solve the equation of motion for a charged fluid element with mass $m$ in this frame, using the
 Lorentz force law and requiring stationary currents:
\be\lab{Lorentz}     
m \frac{d\vec{v'}}{dt} = q \vec{v'}\times \vec{B'} + q \vec{E'} - \mu m \vec{v'} = 0\, ,
\ee
where the last term describes the drag force on a fluid element with mass $m$
on which some external (in this case electromagnetic) force is being exerted,
with $\mu$ being the drag coefficient. 
The nonrelativistic form of (\ref{Lorentz}) is justified by the 
aforementioned assumption  $|\vec{v}|/|\vec{u}|\ll 1$. 
The calculation of $\mu m$ from first principles 
is an interesting open question.  In QCD it may be accessible
via a lattice calculation; in ${\cal N}=4$ supersymmetric Yang-Mills (YM)
theory it should be accessible via a holographic calculation.  At present
its value is known precisely only for heavy quarks in ${\cal N}=4$ SYM theory,
in which~\cite{Herzog:2006gh,CasalderreySolana:2006rq,Gubser:2006bz}
\begin{equation}
\mu m=\frac{\pi \sqrt{\lambda}}{2} T^2\,,
\label{DragCoeff}
\end{equation}
with $\lambda\equiv g^2 N_c$ the 't Hooft coupling, $g$ being the gauge coupling
and $N_c$ the number of colors.  For the purpose of our order-of-magnitude estimate,
we shall use (\ref{DragCoeff}) with $\lambda=6\pi$.  As in our (crude) treatment
of the electric conductivity $\sigma$, and as there for the purpose of 
obtaining our estimates from 
a mostly analytic calculation, we shall also approximate $\mu m$ as a constant.
Throughout this paper we shall choose the constant value of $\mu m$
to be that in (\ref{DragCoeff}) at $T=1.5 T_c$ 	with $T_c\sim 170$~MeV.

In the local fluid rest frame, we look for stationary currents for the up and down quarks and anti-quarks. 
%(Leaving out the strange quarks
%is less of a simplifying assumption than others we have already made.) 
%and we average the velocity for the positively and negatively charged particles separately. 
We assume that the particle density for $u$ and $d$ quarks and antiquarks are all the same, 
thus neglecting any chemical potentials for baryon number or isospin.  (Leaving out the
strange quarks and neglecting any chemical potentials for baryon number or isospin
are  less serious simplifying assumptions than the others that we have already made.)
With these assumptions,  the average velocity for the positively charged species  is $(\vec{v'}_u+ 
\vec{v'}_{\bar{d}})/2$ and
that for the negatively charged species is  $(\vec{v'}_d+ \vec{v'}_{\bar{u}})/2$.   
Having found $\vec{v'}$ for the positively charged particles (and $-\vec{v'}$ for the
negatively charged particles)
we next
transform the four velocity $v'^\mu$ back to the center-of-mass 
frame, obtaining a four velocity,  that we can denote by $V^{+\mu}$ or $V^{-\mu}$, that describes
the sum (in the sense of the relativistic addition of velocities) of $\vec u$ and the
additional charge-dependent velocity $\vec v$ or $-\vec v$.
That is, the four-velocity $V^{+\mu}$ (or $V^{-\mu}$)  
includes both the velocity of the positively (or negatively)
charged particles due to electromagnetic effects and the much
larger, charge-independent, velocity $\vec u$ of the expanding plasma.
Finally, we apply the Cooper-Frye freezeout procedure~\cite{Cooper:1974mv}, 
taking  $V^{+\mu}$ and $V^{-\mu}$
%$u^\mu + v^\mu$ and $u^\mu-v^\mu$ 
as the four-velocity 
for positively and negatively charged particles, integrating over
the freezeout surface, and
calculating the spectra of charged pions and (anti)protons as a function of the transverse momentum
$p_T$, the azimuthal angle in momentum space $\phi_p$ and the 
momentum-space rapidity $Y$.  Integrating the spectra against 
$\cos \phi_p$ yields the directed flow $v_1^+(p_T,Y)$ (and $v_1^-(p_T,Y)$) for positively (and negatively)
charged particles.

After solving Maxwell's equations in Section~\ref{EMT}, in Section~\ref{HydroGubserSection} 
we present our implementation of
Gubser's solution for $\vec u$, including the hadron spectra that result
from it after freezeout.  In Section~\ref{ElectricCurrentSection} we present
the calculation of $\vec v$, and from it $v_1$, that we have just sketched.
We present our estimates of the charge-dependent $v_1$ for 
pions and protons in heavy ion collisions at the LHC and RHIC.   We close in Section~\ref{ObservablesSection} 
with some suggested
correlation observables designed to pull out the effects of the magnetic field whose
magnitude we have estimated, and a look ahead.

%The paper is organized as follows. In the next section we determine the profile of the electromagnetic field in the lab frame by solving Maxwell's equations. Section 3 introduces Gubser's hydrodynamic flow and the Cooper-Frye freez-out analysis to calculate the spectra that comes out of the hydrodynamics. Here we also fix the parroters of Gubser's flow by comparing the spectra to ALICE data. In section 4, we calculate the total fluid 4-velocity including the effects of the charged currents on top of Gubser's fluid velocity. We summarize and discuss our results in the final section.    

\section{Computing the electromagnetic field}
\label{EMT}

 In this Section we determine the electromagnetic field in the center-of-mass 
 frame. 
 % We consider a coordinate system where $\vec{B}$ is directed in the y-axis, whereas z is the beam direction. 
The magnetic field is produced by the charged ions in a non-central collision. 
We begin by considering a single point-like charge located at the position $\vec{x'}_\perp$ in the
transverse plane
moving in the $+z$ direction with velocity $\vec\beta$.
Our coordinates are as in Fig.~\ref{fig1}.
Using Ohm's law  $\vec{J} = \s \vec{E}$  for the current produced in the medium, one finds 
the wave equations
%the following wave-equations for the magnetic and electric fields proceed by a point-like charge moving in the +z-direction with velocity $v$:    
 \bea
\lab{waveB}
\nabla^2\vec{B} -\6^2_t \vec{B} &-&\sigma \6_t \vec{B} = \nn\\{}
&& -e\beta \nabla \times \le[\hat{z} \delta(z-\beta t)\delta(\vec{x}_\perp-\vec{x'}_\perp) \ri] ,\\
\lab{waveE}
\nabla^2\vec{E} -\6^2_t \vec{E} &-&\sigma \6_t \vec{E} = -e\nabla \le[  \delta(z-\beta t)\delta(\vec{x}_\perp-\vec{x'}_\perp)\ri] +\nn\\ {}&& e\beta\hat{z} \6_t\le[  \delta(z-\beta t)\delta(\vec{x}_\perp-\vec{x'}_\perp)\ri]  .
\eea
Solution of these equations is straightforward by the method of Green functions. 
We evaluate the $y$-component of $\vec B$ 
at an arbitrary spacetime point $(t,z,\vec x_\perp)$ in the forward
lightcone, $t>|z|$.  We shall write the spacetime point
in terms of its proper time $\tau\equiv \sqrt{t^2-z^2}$ and spacetime rapidity
$\eta\equiv\arctanh(z/t)$ as well as $x_\perp \equiv |\vec x_\perp|$ and the 
azimuthal angle $\phi$.
%for an arbitrary frame with pseudo-rapidity $\eta$ produced by 
We find that $B_y$ due to a $+$ mover at location 
$\vec x'_\perp$ and $z'=\beta t$ is given by 
\bea\lab{Bfull3}
 e B^+_y(\tau,\eta,\xp,\phi) &=& \alpha\sinh(Y_b)(x_\perp \cos\phi -  x'_\perp \cos\phi') \nn\\
{}&&  \frac{(\frac{\s|\sinh(Y_b)|}{2} \sqrt{\Delta} +1)}{\Delta^{\frac32}} e^A\, ,
 \eea
 where $\alpha=e^2/(4\pi)$ is the electromagnetic
 coupling, 
 $Y_b\equiv\arctanh(\beta)$ is the rapidity of the $+$ mover, and we have defined
 \bea\lab{ADelta}
\!\!\!\!\!\!\!\!\!\! A \!\!&\equiv&  \!\!\! {\frac{\s}{2}\left(\tau\sinh(Y_b)\sinh(Y_b-\eta)-
 |\sinh(Y_b)|\sqrt{\Delta}\right)} \\  
\!\!\!\!\!\!\!\!\!\!  \Delta \!\!&\equiv& \!\!\!\tau^2\sinh^2(Y_b-\eta) + \xp^2+\xp^{'2}\nn\\{}&& - 2\xp\xp'\cos(\phi-\phi')\,  .
\eea
We were able to obtain this analytic solution because we are treating $\sigma$ as constant,
throughout all space and time, with the value $\sigma=0.023$~fm$^{-1}$ chosen
as described in Section~\ref{intro}. 
The finite size and finite duration of the fluid of interest will enter our calculation in Section~\ref{HydroGubserSection},
via the calculation of the freezeout surface. As a check, note  that upon 
setting $\sigma=0$ in (\ref{Bfull3}) one recovers the standard result 
for $B_y$ in vacuum, as in Ref.~\cite{Kharzeev:2007jp}.

A similar calculation shows the $x$-component of the electric field produced by the $+$ moving 
particle is given by
\be\lab{Efull3}
 e E^+_x(\tau,\eta,\xp,\phi) = \,  e B^+_y(\tau,\eta,\xp,\phi) \coth(Y_b-\eta)\, .
 \ee
The other components of the electromagnetic field will turn out to be irrelevant.  

Now we need to evaluate the total $B_y$ and $E_x$ fields produced by 
all the protons in the two colliding nuclei, some of which are spectators, meaning that
at their $\vec{x'}_\perp$ location one finds either $+$ movers or $-$ movers but not both, and others
of which are participants, meaning that at their locations one has both $+$ and $-$ movers.
The spectators have the same rapidity after the collision as they did before it, referred to
as beam rapidity and denoted $Y_0$.  (At RHIC, $Y_0\simeq 5.4$ and at the LHC, $Y_0\simeq 8.$)
Because the participant protons lose some rapidity in the collision, 
after the collision they have some distribution of rapidities $Y_b$.
We shall use the empirical distribution~\cite{Kharzeev:2007jp,Kharzeev:1996sq}
\be\lab{pardist}
f(Y_b) = \frac{a}{2\sinh(a Y_0)}e^{aY_b}, \qquad -Y_0\leq Y_b\leq Y_0\, ,
\ee
for the $+$ moving participants, choosing
$a\approx 1/2$ for both RHIC and LHC collisions. 
(This value of $a$ corresponds to the string junction exchange intercept in Regge
 theory~\cite{Kharzeev:1996sq}
and is consistent with experimental data on baryon stopping~\cite{Kharzeev:1996sq,Abbas:2013rua}.)
After the collision
the $-$ moving participants have 
the same distribution with $Y_b$ replaced by $-Y_b$.
We must then add up the $B_y$ and $E_x$ produced by all the spectators and participants.
Denoting the magnetic field due to spectators and 
participants moving in the +(-) z direction by $\vec{B}^+_s$ ($\vec{B}^-_s$) and $\vec{B}^+_p$($\vec{B}^-_p$), the total magnetic field will be given by $\vec{B} = \vec{B}^+_s +\vec{B}^-_s + \vec{B}^+_p +\vec{B}^-_p$.

\begin{figure}[t!]
 \begin{center}
\includegraphics[scale=0.68]{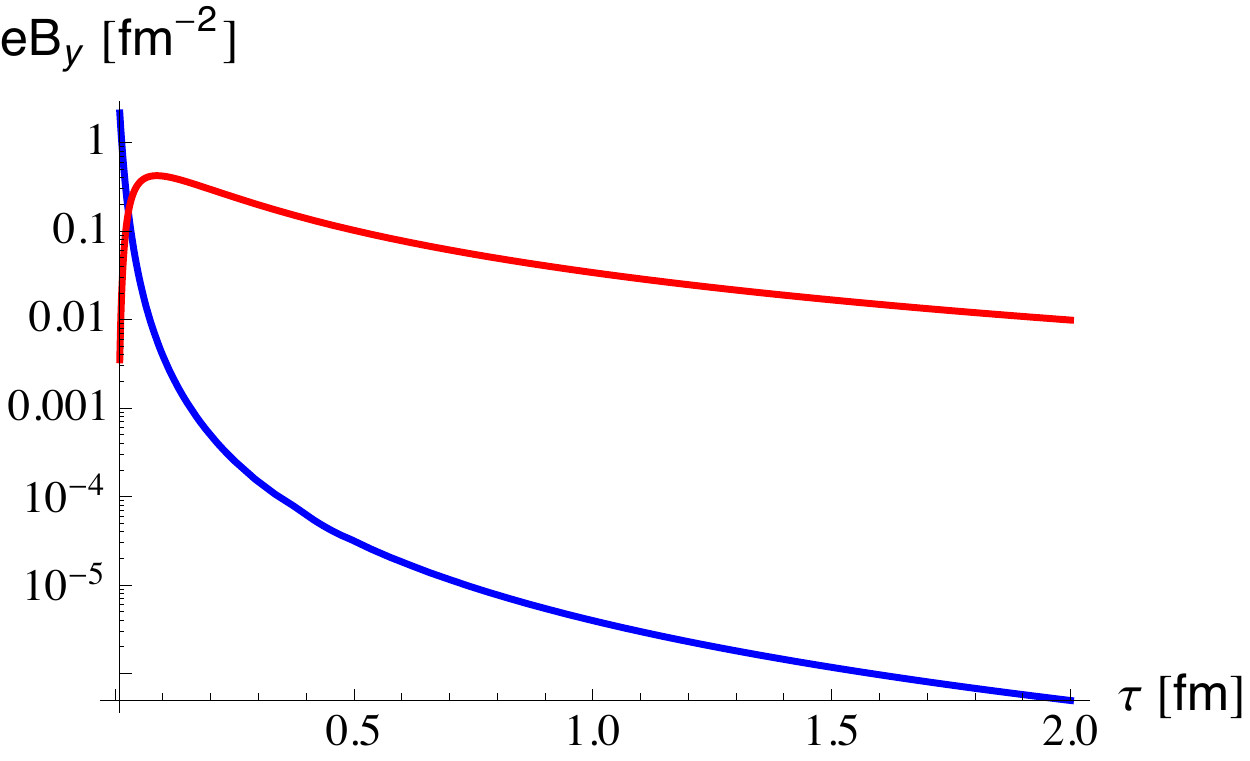}
 \end{center}
 \caption[]{Magnetic field $B_y$ perpendicular to the reaction plane 
 produced by the spectators in a heavy ion collision with impact parameter $b=7$~fm 
 at the LHC. The value of $eB_y$
at the center of the collision, at $\eta=0=x_\perp$, is plotted as a function of $\tau$.  
The blue curve shows how rapidly $B_y$ at $\eta=0=x_\perp$
 would decay as the spectators recede if there were no medium present, i.e.~in vacuum with $\sigma=0$.
  The presence of a conducting
 medium with $\sigma= 0.023$ fm$^{-1}$ substantially delays the
 decay of $B_y$ (red curve). 
 At very early times before any medium has formed, when
 the blue curve is well above the red curve
 the blue curve is a better approximation.  We shall use the red curve
 throughout, though, because our calculation is not sensitive to these 
 earliest times.
 }
\label{figmag}
\end{figure}

Let us first look at the contribution from the spectators.  We shall make the
simplifying assumption that the protons in a nucleus are uniformly distributed
within a sphere of radius $R$, with the centers of the spheres located
at $x=\pm b/2$, $y=0$ and moving along the $+z$ and $-z$ directions
with velocity $\beta$.  We shall take $R=7$~fm and $b=7$~fm.
If we project the probability distribution for the protons in either the $+$ moving
or the $-$ moving nucleus onto the transverse plane it takes the form
\be\lab{rhoHS} 
\rho_\pm(\xp)  = \frac{3}{2\pi R^3}\sqrt{R^2 - \left(\xp^2\pm b\,\xp\cos(\phi)+\frac{b^2}{4}\right)} .
\ee
In a collision with impact parameter $b\neq 0$ the $+$ and $-$ moving spectators
are each located in a crescent-shaped region of the 
$\vec x'_\perp$-plane and
one can write the total electromagnetic field produced by all the spectators as~\cite{Kharzeev:2007jp}
\bea\lab{totBys}
e  B_{y,s} &=&  -Z\int_{-\frac{\pi}{2}}^{\frac{\pi}{2}} d\phi' \int_{x_{\rm in}(\phi')}^{x_{\rm out}(\phi')} d\xp' \xp' \rho_-(\xp') \\
{}&&\hspace{-0.2in}\times\le(eB^+_{y}(\tau,\eta,\xp,\pi-\phi) + eB^+_{y}(\tau,-\eta,\xp,\phi)\ri)\, , \nn\\
\lab{totExs}
e  E_{x,s} &=&  Z\int_{-\frac{\pi}{2}}^{\frac{\pi}{2}} d\phi' \int_{x_{\rm in}(\phi')}^{x_{\rm out}(\phi')} d\xp' \xp' \rho_-(\xp') \\
{}&&\hspace{-0.2in}\times\le(-eE^+_{x}(\tau,\eta,\xp,\pi-\phi) + eE^+_{x}(\tau,-\eta,\xp,\phi)\ri)\, ,\nn 
\eea
with $B_y^+$ and $E_x^+$ defined in (\ref{Bfull3}) and (\ref{Efull3}). 
Here 
%$\theta_{\pm}$ is a projection of the spatial extend of the nuclei moving in the $\pm$ z direction on the transverse plane: 
%\be\lab{thetadef} 
%\theta_\pm(\xp) = \theta[R^2 - |\vec{\xp} \pm \frac{b}{2} \hat{x}|^2] \, ,  
%\ee and 
$x_{\rm in}$ and $x_{\rm out}$ are the endpoints of the $\xp'$ integration 
regions that define the crescent-shaped loci where one finds either $+$ movers
or $-$ movers but not both.  They are given by 
\be\lab{xpm} 
x_{\rm in/out}(\phi') = \mp \frac{b}{2}\cos(\phi') + \sqrt{R^2-\frac{b^2}{4}\sin^2(\phi')}\, .
\ee
We have taken $Z=79$ and $Z=82$ for heavy ion collisions
at RHIC and the LHC, respectively.
In Fig.~\ref{figmag} we plot $eB_y$ produced by the spectators at the 
center of a heavy ion collision at the LHC.  We see that, as other
authors have shown 
previously (see Refs.~\cite{Skokov:2009qp,Tuchin1,Voronyuk:2011jd,Deng:2012pc,Tuchin2,McLerran:2013hla}, 
in particular Fig. 4 in Ref.~\cite{Tuchin2})
the presence of the conducting 
medium delays the decrease in the magnetic field.  This is Faraday's Law in action,
and it tells us that an electric current, indicated schematically
by $J_{\rm Faraday}$ in Fig.~\ref{fig1}, has been induced in the plasma.
Our goal in subsequent Sections will be to estimate the observable
consequences of the presence of such a current.

%%%%%%%%%%%%%%%%%%%%%%%%%%%%%%%%%%%%%%%%%%%%%%%
%%%%%%%%%%%%%%%%%%%%%%%%%%%%%%%%%%%%%%%%%%%%%%%%%%%%%%%%%%%%%%%

%Now, let us turn to the contribution of the participants. We assume that these participants obey the following distribution in rapidity \cite{Kharzeev:2007jp,Kharzeev:1996sq}: 
%\be\lab{pardist}
%f(Y) = \frac{a}{2\sinh(a Y_0)}e^{aY}, \qquad -Y_0\leq Y\leq Y_0\, ,
%\ee
%with $a\approx 1/2$ both for RHIC and LHC. 

A similar calculation to that for the spectators
shows that the total contribution to $B_y$ and $E_x$ from the participants is given by
\bea\lab{totByp}
e  B_{y,p} &=&  -Z\int_{-Y_0}^{Y_0} dY_b f(Y_b) \int_{-\frac{\pi}{2}}^{\frac{\pi}{2}} d\phi' \int_{0}^{x_{\rm in}(\phi')} d\xp' \xp' \rho_-(\xp')\nn\\
{}&& \hspace{-0.25in}\times \le(eB^+_y(\tau,\eta,\xp,\pi-\phi) + eB^+_y(\tau,-\eta,\xp,\phi)\ri), \\
\lab{totExp}
e  E_{x,p} &=&  Z\int_{-Y_0}^{Y_0} dY_b f(Y_b) \int_{-\frac{\pi}{2}}^{\frac{\pi}{2}} d\phi' \int_{0}^{x_{\rm in}(\phi')} d\xp' \xp' \rho_-(\xp')\nn\\ 
{}&&\hspace{-0.25in}\times \le(-eE^+_x(\tau,\eta,\xp,\pi-\phi) + eE^+_x(\tau,-\eta,\xp,\phi)\ri), 
\eea 
where the integration regions have been chosen to correspond to the almond-shaped locus in the
transverse plane where one finds both $+$ and $-$ movers.
Finally, the total electromagnetic field is given by the sum of the contribution of the spectators in  (\ref{totBys}), (\ref{totExs}) and the participants in (\ref{totByp}), (\ref{totExp}). The other components of the electromagnetic field will be irrelevant because $B_z=0$. %Finally we note that, even though the contribution of spectators and participants to the electromagnetic field are of the same order, mai  
In most, but not all, locations in spacetime the contribution of the participant protons to both $B_y$ and $E_x$
is substantially smaller than that of the spectators.  We have checked that eliminating the contribution
from the participants changes the final results that we obtain below for the directed flow by at most 
10\%, typically much less.  

\section{Hydrodynamics and Freezeout}
\label{HydroGubserSection}

%QGP expanding in the beam direction can be described by considering a boost invariant Bjorken flow in the beam direction, that is known to be a good approximation at mid-rapidity. 

As we have already noted in Section~\ref{intro}, we shall use the analytic solution to the
equations of relativistic viscous conformal hydrodynamics found recently by Gubser~\cite{Gubser}
that describes the boost invariant longitudinal expansion and the hydrodynamic
transverse expansion of a circularly symmetric blob of strongly coupled conformal plasma
with four-velocity $u^\mu(\tau,\eta,x_\perp)$, independent of the azimuthal angle $\phi$.
We shall then place this hydrodynamic solution in the electric and magnetic fields
computed  in Section~\ref{EMT}, and determine the small additional
charge-dependent velocity $\vec{v}$ that results.
%
%However, we will also be interested expansion in the transverse directions. Recently an analytic solution has been constructed by Gubser \cite{Gubser} where the plasma also has a non-trivial transverse component. 
%
We refer to  Ref.~\cite{Gubser} for details of Gubser's solution and 
confine ourselves here to a brief summary. 
%is best introduced in a coordinate system  that consists of the proper time $\tau$, rapidity $\eta$, transverse radial direction $\xt$ and the angle on the transverse plane (interaction plane) $\f$:\be\lab{met1} 
%ds^2 = -d\tau^2 + \tau^2 d\eta^2 + d\xt^2 + \xt^2 d\f^2\, .
%\ee    
%Gubser's solution only has two non-trivial components of the 4-velocity, that is 
The only nonzero components of $u^\mu$ are
$u^\tau$, which describes the boost-invariant longitudinal
expansion, and $u^\perp$, which describes the transverse expansion. 
%This translates into all components  non-vanishing in cartesian coordinates, therefore the flow describes an expanding fluid both in the beam axis and in the transverse directions. 
They are given by~\cite{Gubser}
\be\lab{Flow} 
u^\tau= {1 + q^2 \tau^2 + q^2 x_\perp^2 \over 2q\tau \sqrt{1+g^2}} \, ,
    \qquad
  u^\perp = {q x_\perp \over \sqrt{1+g^2}} \,,
 \ee 
where %only the non-zero components of $u^\mu$ in the $(\tau,\eta,x_\perp,\phi)$ coordinate system are shown and we defined
 \be\lab{gdef}
  g \equiv {1 + q^2 x_\perp^2 - q^2 \tau^2 \over 2 q \tau} \, . 
 \ee
The fluid four-velocity $u^\mu$ in the solution 
is specified by a single parameter denoted by $q$,
with the dimension of 1/length. ($q$ is unrelated to charge.)
The transverse size of the plasma is proportional to $1/q$.
%changing $q$ also changes the initial temperature of
%the plasma.
The local  temperature of the plasma 
is then given by~\cite{Gubser}
 \bea\lab{TG}
  T &=& {1 \over \tau f_*^{1/4}} \bigg( {\hat{T}_0 \over (1+g^2)^{1/3}} + {{\rm H}_0\, g \over \sqrt{1+g^2}} \nn\\ 
  {} &\times& 
   \left[ 
      1 - (1+g^2)^{1/6} {}_2 F_1\left( {1 \over 2}, {1 \over 6}; {3 \over 2}; -g^2
        \right) \right] \bigg) \,,
 \eea
where the first the term, proportional to the dimensionless 
parameter $\hat T_0$, corresponds to an ideal fluid and the second term incorporates dissipative effects 
due to the shear viscosity $\eta$.  
The initial temperature of the plasma is proportional to the parameter $\hat T_0$,
and is also affected by the choice of the parameter $q$.
The expression (\ref{TG}) introduces two further dimensionless parameters that we shall
choose as in Ref.~\cite{Gubser}. $f_*$ is the parameter that relates the energy density
of the plasma $\varepsilon$ to the local temperature, $\varepsilon=f_* \,T^4$, and we shall
choose the value $f_*=11$, reasonable for the QCD quark-gluon plasma with
$T\sim 300$~MeV~\cite{Borsanyi:2010cj}.
$H_0$ is the parameter that controls the strength of viscous corrections; it
is defined by $\eta=H_0\, \varepsilon^{3/4}$.
We
choose the value $H_0=0.33$ that corresponds to $\eta/s=0.134$, as has
been estimated for $SU(3)$ gluodynamics~\cite{Meyer}.
The local energy density $\varepsilon$ and the fluid four-velocity $u^\mu$ fully specify
the energy-momentum tensor of the fluid.

%$\hat{T}_0$ is a parameter that specifies the initial temperature of the plasma and $f_*$ is the ratio of the energy density to temperature $f_* = \epsilon/T^4$. $H_0$ is a parameter that determines that the strength of viscous corrections.    

%, and the stress tensor $T_{\mu\nu}$, in particular the energy density and the local temperature of the fluid, see top figure in figure \ref{fig2}. First order viscous corrections are included in the flow. 

%\footnote{There are actually two more parameters for which we choose the same values as  $f_*=11$ and $H_0=0.33$ as in \cite{Gubser}.}

%%%%%%%%%%%%%%%%%%%%%%%%%%%%%%%%%%%%%%%%%%%%%%%
%\begin{figure}[h!]
% \begin{center}
%\includegraphics[scale=0.58]{TGubser}
%\includegraphics[scale=0.6]{Tfreez}
% \end{center}
% \caption[]{ Parametric plot of isothermal curves that results from Gubser's solution for a choice of parameters $%\hat{T}_0=10.8$ at fixed $q=1/6.4$ fm$^{-1}$. }
%\label{TGubser}
%\end{figure}
%%%%%%%%%%%%%%%%%%%%%%%%%%%%%%%%%%%%%%%%%%%%%%%%%%%%%%%%%%%%%%%
 %Gubser's solution has 4 parameters $q$, $\hat{T}_0$, $f_*$ and $H_0$. For $f_*$ we take the standard value   $f_* = \epsilon/T^4=11$. $H_0$ can be determined by the lattice value \cite{Meyer} for the shear viscosity $\eta/s=0.134$ for $SU(3)$ gluodynamics as \cite{Gubser} $H_0 = 0.33$.

It remains to fix the parameters $q$ and $\hat{T}_0$.  
Together they determine the initial temperature profile of the plasma
at some fiducial early time that should be comparable to or greater
than the time at which a hydrodynamic description becomes valid.
Hydrodynamic calculations appropriate for heavy ion collisions 
at the LHC, for example those in Refs.~\cite{Shen:2012vn},
suggest that at $\tau=0.6$~fm the initial temperature should be between 445 MeV 
and 485 MeV.   It is not possible to use this initial temperature to
fix $q$ or $\hat{T}_0$, however, because the  the temperature
profile as a function of $x_\perp$ is quite different in Gubser's
solution than in a heavy ion collision: in Gubser's solution the temperature profile
is both more peaked at $x_\perp=0$ and has a heavier large-$x_\perp$ tail relative to a Woods-Saxon
distribution with its flat middle and damped tails.
The parameters $q$ and $\hat{T}_0$ also implicitly determine
the radial velocity profile at the end of the hydrodynamic evolution,
which in turn determines the hadron spectra after freezeout.
%
%On the other hand, both of these parameters substantially affect the hadron spectra obtained from Gubser's hydrodynamics after freeze-out. We find  that these two affects are in competition with each other and one has to optimize the choice of parameters so that both the initial temperature is in accord with existing hydro simulations and the spectra is in accord with the data at LHC. 
%
Our approach, therefore, is to explore the two parameter space
looking for values that give reasonable final state spectra to mock up
heavy ion collisions at the LHC in the 20 - 30\% centrality class (i.e.~the collisions in
the 20th-30th percentile in impact parameter, which have impact parameters around $7 - 8$ fm.
We have found that choosing
$\hat{T}_0=10.8$ and $q^{-1}=6.4$~fm yields reasonable pion and proton spectra, as 
we shall show below.
This choice yields a temperature of 617~MeV at the
center of the collision at $\tau=0.6$~fm and an average
temperature within $x_\perp < 7$~fm at $\tau=0.6$~fm of 458~MeV.
For heavy ion collisions at RHIC we find instead that choosing
$\hat{T}_0=7.5$ and $q^{-1}=5.3$~fm yields reasonable pion and proton spectra.
With this choice, at $\tau=0.6$~fm the temperature at $x_\perp=0$ is 488~MeV
and the average temperature within $x_\perp<7$~fm is 326~MeV.

%WE MAKE THE FOLLOWING CHOICES.

We calculate the hadron spectra for the pions and the protons by applying the Cooper-Frye 
freezeout procedure to Gubser's hydrodynamic solution.
The hadron spectrum for particles of species $i$ with mass $m_i$ will depend on transverse
momentum $p_T$, momentum space rapidity $Y$ and the azimuthal angle
in momentum space $\phi_p$.  These are related to $p^\mu$ by
%$p^0$, $p^x$, $p^y$ and $p^z$ by
\begin{eqnarray}
p^0 &=& m_T\cosh Y,\nn\\
p^z &=& m_T\sinh Y, \nn\\
p^y &=& p_T \sin\f_p,\nn\\
p^x &=& p_T\cos\f_p,
\end{eqnarray} 
where we have defined the transverse mass
$m_T\equiv \sqrt{p_T^2 + m_i^2}$. 
To establish notation, note that the dependence of the hadron spectrum
on $\phi_p$ can be expanded as
\bea\lab{flowpars}
S_i &\equiv& p^0 \frac{d^3N_i}{dp^3}=\frac{d^3N_i}{p_T dY dp_T d\f_p } \\ 
{} &=&  v_0 \le(1+ 2\,v_1\cos(\f_p-\pi) + 2\,v_2 \cos2\f_p +\cdots \ri),  \nn
\eea
where in general the parameters $v_n$ will depend on $Y$ and $p_T$.
Note that the sign of $v_1$ is conventionally defined such that if the spectators moving toward
positive $z$, i.e.~moving with positive $Y$, were deflected away from the center of
the collision that would correspond to a positive $v_1$.  We see in Fig.~\ref{fig1} 
that, with our choices of conventions, the spectators moving toward positive $z$ are
at negative $x$.  This means that for us 
$v_1>0$ corresponds to directed flow
toward negative $x$, as we have already indicated in the labelling of Fig.~\ref{fig1}.
This is why $v_1$ multiplies $\cos(\f_p-\pi)$, not $\cos\f_p$, in (\ref{flowpars}).

%
%Gubser's solution describes a boost invariant flow that is centrally symmetric. Thus the only non-vanishing flow parameter is $v_0$ that can only depend on $p_T$. The standard 4-momentum variables $p_T$, $p_0$, $Y$ and $\f_p$ are defined in terms of the momenta in the Minkowski frame (\ref{met1}) as 
%
%Transforming these expressions to the 4-momenta in the Bjorken frame (\ref{met1}) one finds, 
%\bea\lab{mom2} 
%p^\tau &=& m_T\cosh(Y-\eta), \,\,\,\, p^\eta = \frac{m_T}{\tau}\sinh(Y-\eta), \nn\\ 
%p^\f &=& \frac{p_T}{\xp} \sin(\f_p-\f), \,\,\,\, p^\perp = p_T\cos(\f_p-\f)\, .
%\eea
%

\begin{figure}[t]
% \begin{center}
\includegraphics[scale=0.65]{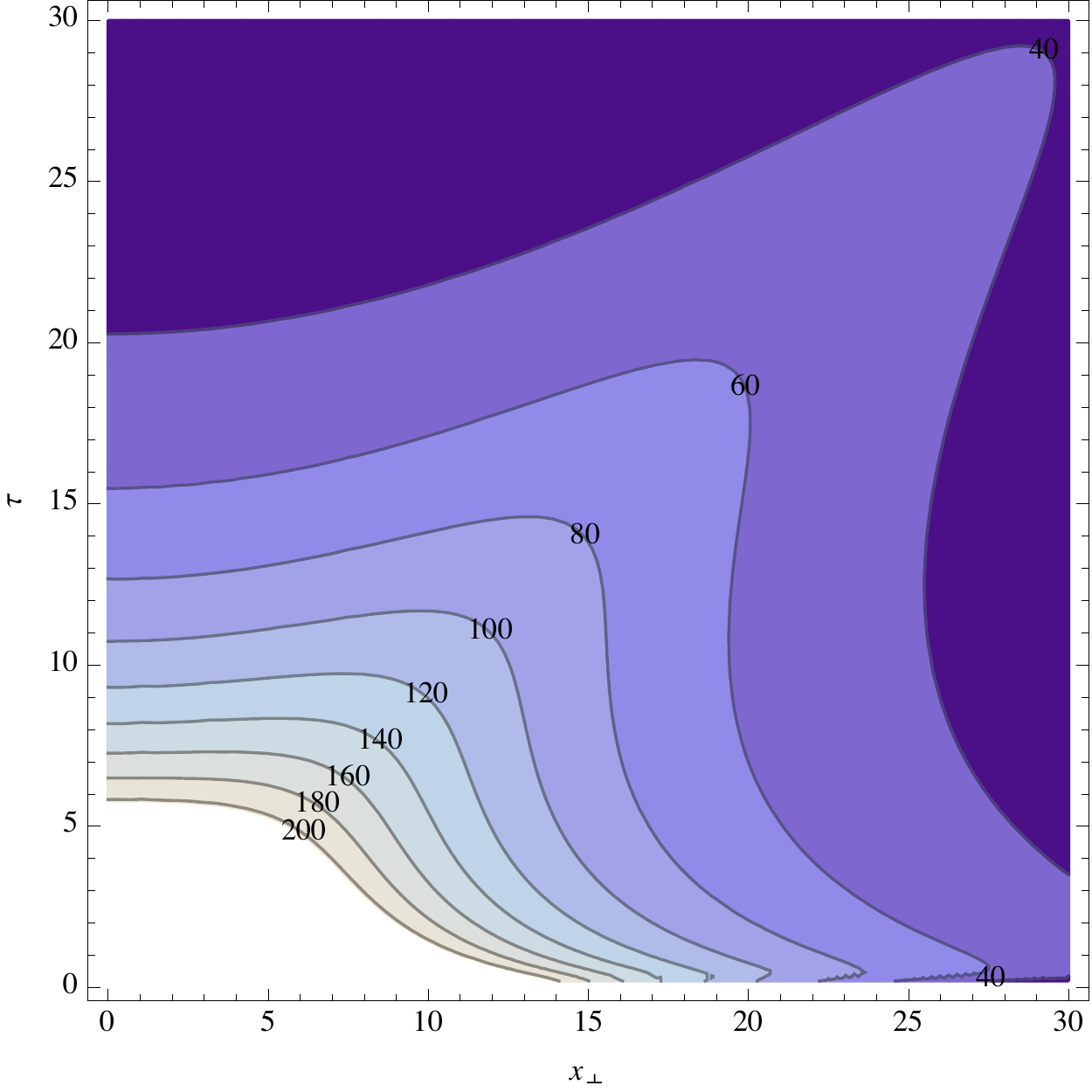}
\includegraphics[scale=.7]{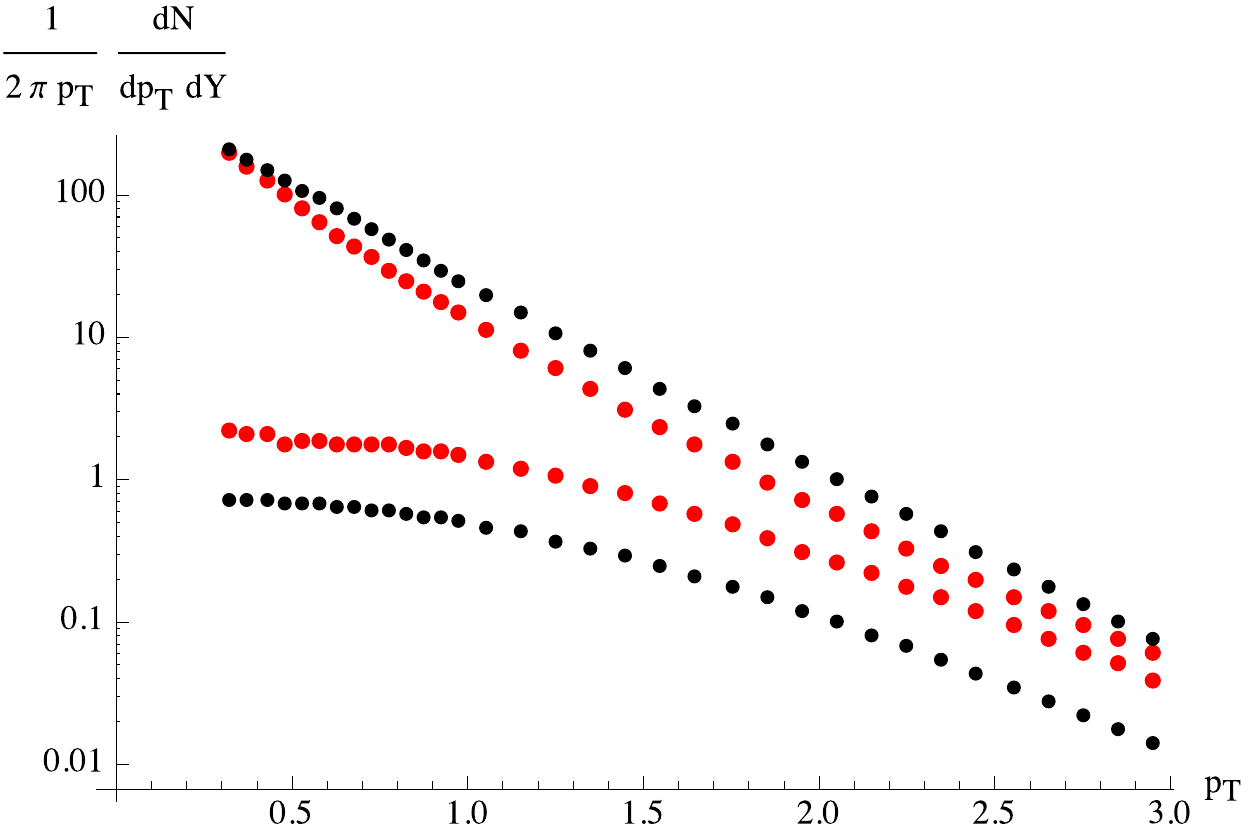}
 \caption[]{Features of Gubser's flow. Top figure illustrates the isothermal curves in the $(x_\perp,\tau)$ plane 
 for Gubser's hydrodynamic solution with the parameters $\hat{T}_0=10.8$ and $q^{-1}=6.4$~fm. 
 We choose the $T=130$~MeV isotherm as the freezeout surface.
 Bottom figure is the comparison of the spectrum of positively charged pions (black, top) and protons (black, bottom) as a function of transverse momentum $p_T$  resulting from Gubser's hydrodynamic solution to the 
spectra for pions (red, top) and protons (red, bottom) in LHC heavy ion collisions with
20-30\% centrality
measured by the ALICE collaboration, as in 
Ref.~\cite{Alice}.
Because we have no chemical potential for baryon number or isospin in our calculation, the spectra of antiprotons and
protons are identical as are the spectra of the negatively and
positively charged pions.}
\label{GubserFlowFigure}
\end{figure}
%%%%%%%%%%%%%%%%%%%%%%%%%%%%%%%%%%%%%%%%%%%%%%%%%%%%%%%%%%%

Gubser's solution is boost invariant and azimuthally symmetric, meaning that it is independent
of $Y$ and $\phi_p$.  In this case, the only nonvanishing $v_n$ is $v_0$, 
and $v_0=(2\pi p_T)^{-1}d^2N/dYdp_T$
depends only
on $p_T$.
We want to calculate $v_0$
%= p^0 dN_i/dp^3$ 
for pions and protons. 
%and compare with ALICE data \cite{Alice}.  
The standard prescription to obtain the hadron spectra from a hydrodynamic flow, assuming
sudden freezeout when the fluid cools to a specified freezeout temperature $T_f$, was
developed by Cooper and Frye~\cite{Cooper:1974mv}.   
We shall take $T_f=130$~MeV for heavy ion collisions at both the LHC and RHIC.
The freezeout surface is the isothermal surface in spacetime at which the temperature 
of Gubser's hydrodynamic solution, given in (\ref{TG}), satisfies
$T(\xp,\tau) = T_f$.
The spectrum for hadrons of species $i$ is then given by~\cite{Cooper:1974mv}  
\be\lab{CF} 
S_i = p^0\frac{d^3N_i}{dp^3} = -\frac{g_i}{(2\pi)^3} \int d\Sigma_\mu\, p^\mu F\le( -\frac{p^\mu u_\mu}{T_f} \ri)\, ,
\ee
where $d\Sigma_\mu$ is the area element on the freezeout surface, 
$u^\mu$ is the 4-velocity of the fluid, $g_i$ is the degeneracy of hadron species $i$ 
and $F(x)$ is a distribution function 
that  we will take as the Boltzmann distribution $F(x) = \exp(-x)$.
As with many of our other simplifying assumptions, we choose Boltzmann rather than
Fermi-Dirac or Bose-Einstein in order to obtain a calculation that
can be done mostly analytically.  (The sign in the argument of $F$ in (\ref{CF}) 
comes from our use of the mostly $+$ signature metric.)
The freezeout surface %in the Bjorken coordinate frame 
is $\Sigma^\mu =(\tau_f(\xp), \eta, \xp,\f)$ where $\tau_f(\xp)$ is the solution of the 
equation $T(\xp,\tau_f) = T_f$.  See Fig.~\ref{GubserFlowFigure} (top). 
The area element perpendicular to the freezeout surface is 
\bea\lab{ael} 
d\Sigma_\m &=& -\epsilon_{\m\n\l\r} \frac{\6\Sigma^\n}{\6\eta} \frac{\6\Sigma^\l}{\6\xp}\frac{\6\Sigma^\r}{\6\f}\sqrt{-g}\, d
\eta\, d\xp d\f \nn\\
{}&=&\le(-1,0,-R_f,0\ri) \xp \tau_f\, d\eta\, d\xp d\f\, ,
\eea
where $\sqrt{-g}=\xp\tau$ 
on the freezeout surface and where we have used the fact that
$dT= (\6T/\6\xp) d\xp +(\6T/\6\tau) d\tau =0$ on the freezeout surface to define
\be\lab{Rf} 
R_f\equiv  - \frac{\6\tau}{\6\xp} =  \frac{\6T}{\6\xp}\Big/\frac{\6T}{\6\tau}\Big|_{T_f}\ .
\ee
This completes the specification of the quantities appearing in the expression (\ref{CF})
for the hadron spectra.

We now calculate (\ref{CF}) for Gubser's flow, for example with the parameters chosen with LHC
heavy ion collisions in mind as we described
above.
% where $V^\mu = u^\mu$ and $u^\mu$ is given in \cite{Gubser}. 
For $u^\mu$ as in Gubser's flow, the argument of the function $F$ simplifies as 
\be\lab{expGubser} 
p^\mu u_\mu = -m_T\, u^\tau \cosh(Y-\eta) + p_T\, u^\perp \cos(\f_p - \f)\, . 
\ee
One can then perform the $\eta$ and $\f$ integrals in (\ref{CF}) analytically, obtaining
\bea 
\hspace{-0.2in}p^0 \frac{d^3N_i}{dp^3}\Bigg|_{G}  &=&   \frac{g_i}{2\pi^2}\int d\xp \xp\, \tau_f(x_\perp) \nn\\
{}&\times&\hspace{-0.06in}\bigg\{m_T\,K_1\left(\frac{m_T u^\tau}{T_f}\right)I_0\left(\frac{p_T u^\perp}{T_f}\right)\nn\\ 
\lab{v0G}
{}&+& \hspace{-0.06in}R_f p_T K_0\left(\frac{m_T u^\tau}{T_f}\right)I_1\left(\frac{p_T u^\perp}{T_f}\right)\bigg\}
\eea  
where $R_f$ was defined in (\ref{Rf}).
% = (\6T/\6\xp)/(\6T/\6\tau)$ evaluated at $T=T_f$.  
As expected, the result is independent of $Y$ and $\f_p$ and only depends on $p_T$. 
One then evaluates the $\xp$ integral on the freezeout surface numerically and obtains the 
spectra of hadrons freezing out from Gubser's hydrodynamic flow as a function of $p_T$.
The results for the charged pion and proton spectra 
%for the choice of parameters described in the beginning of this section, 
are presented in Fig.~\ref{GubserFlowFigure} (bottom). 

We observe that Gubser's flow with this choice of parameters does not yield fully satisfactory spectra --- in particular there are too few protons relative to pions --- but at a qualitative level it reproduces many features of the
spectra in LHC heavy ion collisions with 20-30\% centrality
measured using the ALICE detector~\cite{Alice}.   
The shortfall in the number of protons comes because we are using a single freezeout temperature $T_f$ instead
of letting the number of each hadron species, for example protons, freezeout first at a somewhat higher 
chemical freezeout temperature or using a hadron cascade
code between $T_c$ and $T_f$.  By assuming thermal and chemical
equilibrium and using hydrodynamics
all the way down to a single freezeout temperature $T_f=130$~MeV 
the proton multiplicity in the final state is being
overly suppressed by the Boltzmann factor at $T=T_f$.  
We see, though, that the shape of the proton spectrum
is reproduced well. 
If we were to use a slightly lower $T_f$, say 120 or 110 MeV, we could improve
the shape of the pion spectrum at the expense of suppressing the proton multiplicity even more
than in Fig.~\ref{GubserFlowFigure}.
Given the simplicity, and the unphysical initial temperature profile, of Gubser's analytic hydrodynamic solution 
and given the crude
freezeout at a single $T_f$ that we are employing, we find it impressive that it is possible
to obtain spectra as reasonable as those in Fig.~\ref{GubserFlowFigure}.

We have also done the exercise of comparing spectra obtained at freezeout from Gubser's hydrodynamic
solution with varying values of $q$ and $\hat{T}_0$ to pion and proton
spectra for  20-30\% centrality heavy
ion collisions at RHIC~\cite{Adler:2003cb}, finding reasonable spectra upon choosing
$q^{-1}=5.3$~fm and $\hat{T}_0=7.5$, values of the parameters
that we already quoted earlier in this Section.  

In the next Section, after we have determined the charge-dependent velocity corresponding
to the electric current we shall re-evaluate (\ref{CF}) upon replacing $u^\mu$ by 
 $V^{+\mu}$  or $V^{-\mu}$
for positively or negatively charged hadrons.

%It is  not surprising that this simple 
%hydrodynamic model could not describe all experimental data.     

%This qualitative comparison becomes also quantitatively satisfactory for the choice  $\hat{T}_0=15$ and $q=1/11$ fm$^{-1}$. In the rest of the paper we shall use these particular values for the Gubser's parameters.  
%However it yields a qualitatively similar shape also in this case, in particular one can clearly see the flattening of the proton spectra for small values of $p_T$. One could use a different set of parameters that would make the comparison with protons better with the expense of worsening the comparison for pions. We prefer not to do this and simply accept the fact that the hydrodynamics is not very suitable for protons. In any case,\section{Charged currents and $v_1$}

%\subsection{Adding the charge current to Gubser's flow}
%\lab{Addition}

\section{Electric current and charge-dependent directed Flow}
\label{ElectricCurrentSection}

We are now ready to study the effects of the magnetic field on the directed flow $v_1$, which 
is the purpose of this paper.
In the center-of-mass frame, 
the magnetic and electric fields are given by the sum of (\ref{totBys}) and (\ref{totByp}), and  (\ref{totExs}) and (\ref{totExp}).  
%The velocity is then given by $\vec{u}_i = u^i/u^0$. 
The fluid velocity in the absence of any electromagnetic effects
is given by $u^\mu$ in Gubser's solution, (\ref{Flow}).   In order 
to obtain the fluid velocity $V^\mu$, including 
electromagnetic effects, at a given spacetime point we first Lorentz
transform by $\Lambda{(-\vec u)}$
to the local fluid rest frame in which $\vec{u'}=0$ at that point
and then use the Lorentz transformed electromagnetic fields in the stationary
current condition (\ref{Lorentz}),
setting $q=+2e/3$ to obtain $\vec{v'}_u$, setting $q=+e/3$ to obtain $\vec{v'}_{\bar d}$,
and averaging these to obtain $\vec{v'}$.  The average drift velocity for negatively charged
particles in the local fluid rest frame 
is obtained similarly and is given by $-\vec{v'}$.
We then Lorentz transform by $\Lambda(\vec{u})$ back to the center-of-mass 
frame, obtaining the total velocity $V^{+\mu}$ and $V^{-\mu}$ for positively and negatively charged
particles via Lorentz transforming $\vec{v'}$ and $-\vec{v'}$ back to the center-of-mass
frame, respectively.

At this point we checked whether our assumption $|\vec{v}|/|\vec{u}|\ll 1$ is indeed satisfied. In order to characterize this assumption in a Lorentz invariant fashion, 
we can calculate the difference between the Lorentz factor for the total velocity $V^{\pm\mu}$, 
including Gubser's $u^\mu$ 
and the excess velocity due to magnetic effects, and the Lorentz factor for $u^\mu$ alone. 
The difference between these turns out to be very small, of order 0.001 or smaller
everywhere  in the $(\eta,\xp,\phi)$ space.
 % \cite{upcoming}. 
 %This is presented in figure \ref{figgamma} and one indeed observes that the difference between the two - that is the contribution to the fluid velocity solely from magnetic effects is indeed tiny.  

%{\bf \large discussion of the ratio's of V and u here.}  
 
Once we have obtained the total velocity $V^{\pm\mu}$ we can use the freezeout procedure described 
in the previous Section to calculate the hadron spectra including electromagnetic effects by
replacing $u^\mu$ in (\ref{CF}) by $V^{+\mu}$ when evaluating the $\pi^+$ and proton
spectra and by $V^{-\mu}$ when evaluating the $\pi^-$ and antiproton spectra.
%
%
%In particular one has to use the total velocity $V$ in the expression   (\ref{CF}).  
%
%We assume that the velocity $u$ of the expanding medium is much larger than the velocity of charged particles  due to charge current $J$.  With this assumption, it is consistent to ignore the back-reaction of $J$ on the expanding medium, hence it is enough to just add the two velocities, the 3-velocity $\vec{u}'$ due to expansion of the medium and the 3-velocity $\vec{v}$ due to the charge current $J$. 
% 
% 
% We already discussed the spectra of pions and protons that follow from Gubser's flow in section \ref{Gubser}. As all other flow parameters $v_i$, $i>0$ vanish for Gubser's flow alone, the spectra shown in figure \ref{fig2} also gives the $v_0$ coefficient in this case. On the other hand  
The change to the $\phi$-integrated $dN/dp_T$, i.e.~the change to $v_0$ defined in (\ref{flowpars}),
that results from using $V^{\pm\mu}$ instead of $u^\mu$ is minuscule, and for all practical purposes it
is fine to use results for $v_0$ obtained as in Section~\ref{HydroGubserSection}.
Because the magnetic field induces an electric current that circulates in the $(x,z)$
plane, see Fig.~\ref{fig1},  when we use $V^{+\mu}$ or $V^{-\mu}$ 
in (\ref{CF}) we obtain a small, but nonzero, directed flow $v_1$ that
is opposite in sign for positively and negatively charged particles.  
Teasing out this charge-dependent $v_1$ is the goal of this paper.
%
%
%it is straightforward to check that contribution of the circular current to $v_0$ is quite small, in accordance with our working assumption that one can treat the circular flow as a perturbation on top of the expanding medium. Our main interest in this paper is $v_1$ that is obviously only due to the circular currents in the plasma and vanishes in the absence of such currents. 
%Therefore, for the purpose of calculating $v_1$ one can safely approximate the total $v_0$ as $v_0$ of expanding  medium only that is presented in figure  \ref{fig2}: 
 %\be\lab{v0tot} 
 %v_0^{tot} \approx v_0\Big|_{Gubser}\, .
 %\ee 
 %
 %We finally	 turn to the question of determining $v_1$. 
 From its definition in  (\ref{flowpars}) we see that $v_1$ is given by 
 \be\lab{v1def}
 v_1(p_T,Y) = \frac{\int_{-\pi}^{\pi} d\f_p\, \cos(\f_p-\pi)\, S_i(p_T,Y,\f_p)}{2\pi v_0}\, .%\equiv \frac{N_i}{2\pi v_0} \, .
 \ee 
Recall from Fig.~\ref{fig1} that our conventions are such that a positive $v_1$ corresponds
to directed flow in the negative $x$ direction.
In evaluating the denominator in (\ref{v1def}) we shall use $v_0$ obtained from $u^\mu$ as in Section~\ref{HydroGubserSection}.
% We now note that the contribution of the velocity $\vec{v'}$ due to B in the total $v_0$ should be negligible if $|\vec{v'}|/|\vec{u}|\ll 1$ and we can replace $v_0$ with the $v_0$ only due to Gubser's flow, $v_0 \approx S_i\big|_G$, equation (\ref{v0G}).  On the other hand, 
There are four integrals to be evaluated in the numerator of (\ref{v1def}), namely integrals
over $\xp$, $\eta$, $\f$ and $\f_p$. 
It turns out that one can evaluate the $\f_p$ integral analytically in terms of Bessel and hypergeometric functions:  
\bea\lab{I1} 
\int_{-\pi}^{\pi} d\f_p &&\hspace{-0.25in} \cos\f_p\, S_i(p_T,Y,\f_p) =  \frac{g_i}{(2\pi)^2}\!\!\int d\eta \,d\xp \,d\f \,\,\xp \,\tau_f(x_\perp) \nn\\
&\times&
e^{-\frac{m_T}{T_f}\left[V^\tau\cosh(Y-\eta) -V^\eta\tau_f \sinh(Y-\eta)\right]}\nn\\
& \times& \Bigg\{ \le(V^\perp\cos\f-\xp V^\f\sin\f\ri) \nn\\
{}& \times &\Bigg[ \frac{m_T\cosh(Y-\eta)}{\sqrt{W}} I_1\le(\frac{p_T}{T_f}\sqrt{W}\ri)   \nn\\
{}& & + R_f\, p_T \frac{v^\perp}{W}\le( I_0\le(\frac{p_T}{T_f}\sqrt{W}\ri)-\Psi_2\le(\frac{p_T^2}{4T_f^2}W\ri)\ri)\Bigg] \nn\\ 
{}& &\quad + \half R_f\, p_T \cos\f \,\Psi_2\le(\frac{p_T^2}{4T_f^2}W\ri)\Bigg\}\, ,
\eea
where we have defined $W \equiv (V^\perp)^2 + \xp^2(V^\f)^2$. 
The three remaining integrals in (\ref{I1}) have to be done numerically. 
After doing so we obtain $v_1(Y,p_T)$ from  (\ref{v1def}).

%as $v_1 = I_1/2\pi v_0$ by substituting (\ref{I1}) and the result above  for $v_0$.

\begin{figure}[t]
 \begin{center}
\includegraphics[scale=.68]{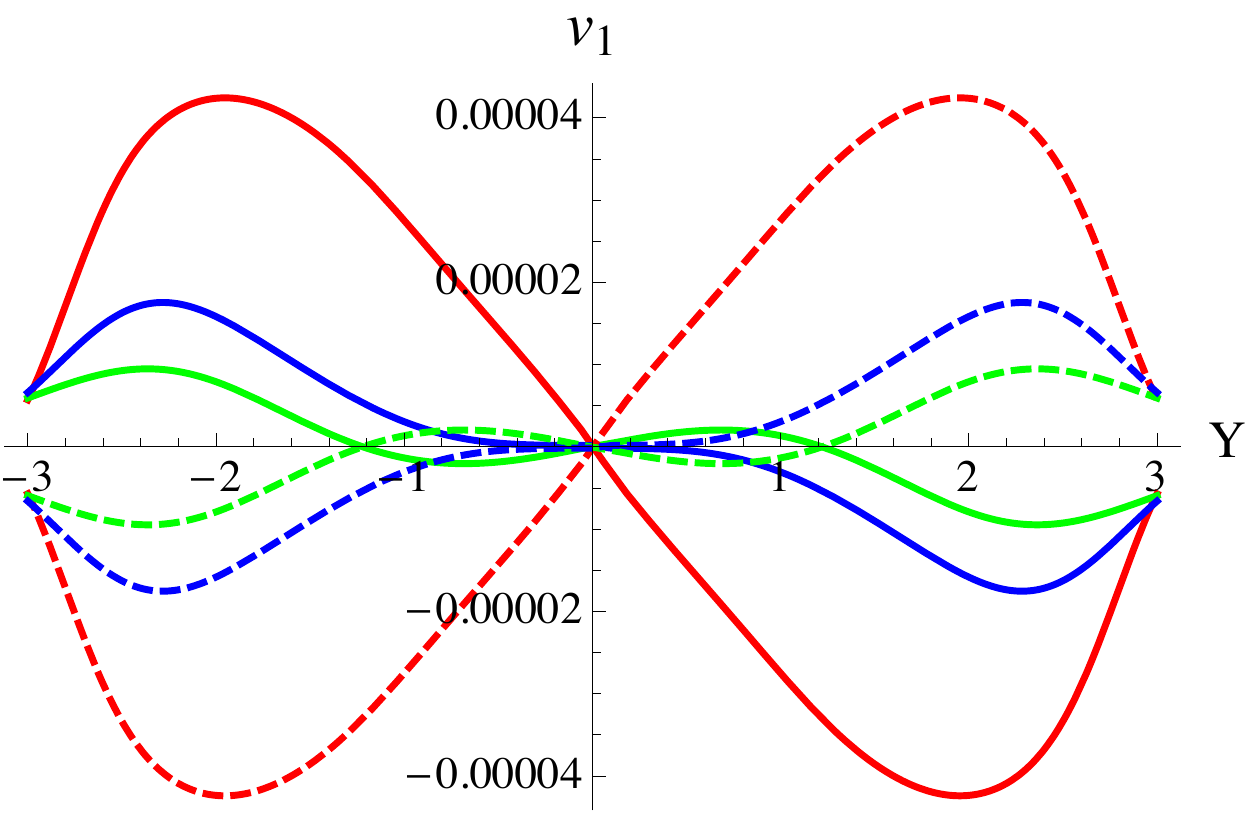}
 \end{center}
\vspace{-.3cm}
 \caption[]{Directed flow $v_1$ for positively charged pions (solid curves) and negatively charged pions (dashed curves) 
in our calculation with parameters chosen 
to give a reasonable facsimile of 20-30\% centrality heavy ion collisions
at the LHC. 
We plot our results for $v_1$ as functions 
of momentum-space rapidity $Y$ at $p_T = 0.25$ (green), 0.5 (blue) and 1 GeV (red). Here
and in all subsequent figures we are only plotting the charge-dependent contribution to
the directed flow $v_1$ 
that originates from the presence of the magnetic field in the collision and that is caused by the Faraday and Hall
effects. This charge-dependent contribution to $v_1$ must be added to the, presumably larger,
charge-independent $v_1$.  For example, if the charge-independent $v_1$ for pions with $Y<0$ and 
$p_T=1$~GeV 
is positive then in that kinematic regime our results correspond to a positive $v_1$ for both $\pi^+$ and
$\pi^-$, with $v_1(\pi^+) > v_1(\pi^-)$ .}
\label{fig5}
\end{figure}

 Figure \ref{fig5} shows $v_1$ for positively and negatively 
 charged pions as  a function of momentum-space rapidity $Y$ at transverse momenta $p_T = 0.5$, 1, and 2 GeV.  
 In this Figure we have chosen the initial magnetic field
 created by the spectators with beam rapidity $\pm Y_0=\pm 8$ and the  participants,
 we have set the parameters specifying Gubser's hydrodynamic solution to
 $\hat T_0=10.8$ and $q^{-1}=6.4$~fm,  we have chosen the electric conductivity 
 $\sigma= 0.023$~fm$^{-1}$ 
 and the drag parameter $\mu m$  in (\ref{Lorentz}) as in (\ref{DragCoeff}) with $T=255$~MeV,
 and we have set the freezeout temperature to $T_f=130$~MeV. 
 As we have described
 in previous Sections, these parameters have been chosen to give a reasonable characterization
 of $v_1$ in 20-30\% centrality heavy ion collisions at the LHC.
Note that here and in the following we will only look at the directed flow
 at values of $|Y|$ that are well below $Y_0$. This is because the trajectories of final-state
 hadrons produced near beam rapidity can be affected by Coulomb interactions
 with the charged spectators at very late times~\cite{Rybicki:2013qla}, long after freezeout,
 and we are neglecting these effects.

%{\bf\large Hall vs. Faraday?}
%In this figure we also include the $v_1$ that would result only by including the Faraday effect. The latter can be obtained by  setting $E'=E$ and omitting the $v'\times B$ term in equation  (\ref{Newton1}). We observe that the Hall effect is in the opposite direction as the Faraday effect and results in a substantial reduction in the directed flow.  
%\vspace{-.3cm}
 %%%%%%%%%%%%%%%%%%%%%%%%%%%%%%%%%%%%%%%%%%%%%%%

%%%%%%%%%%%%%%%%%%%%%%%%%%%%%%%%%%%%%%%%%%%%%%%%%%%%%%%%%%%
%Figure \ref{fig6} shows $v_1$ for positively charged pions as  a function of transverse momentum at fixed  $Y=1$ and 2.   
%\vspace{-.3cm}
 %%%%%%%%%%%%%%%%%%%%%%%%%%%%%%%%%%%%%%%%%%%%%%%
%\begin{figure}[h!]
% \begin{center}
%\includegraphics[scale=.85]{v1pT-LHC}
%\includegraphics[scale=.85]{v1-for-pT-09GeV_sum_method}
% \end{center}
%\vspace{-.3cm}
% \caption[]{$v_1$ for positively charged pions as a function of transverse momentum at  $Y=1$ (red) and 2 (blue). }
%\label{fig6}
%\end{figure}
%%%%%%%%%%%%%%%%%%%%%%%%%%%%%%%%%%%%%%%%%%%%%%%%%%%%%%%%%%%

%%%%%%%%%%%%%%%%%%%%%%%%%%%%%%%%%%%%%%%%%%%%%%%
\begin{figure}[t]
 \begin{center}
\includegraphics[scale=.68]{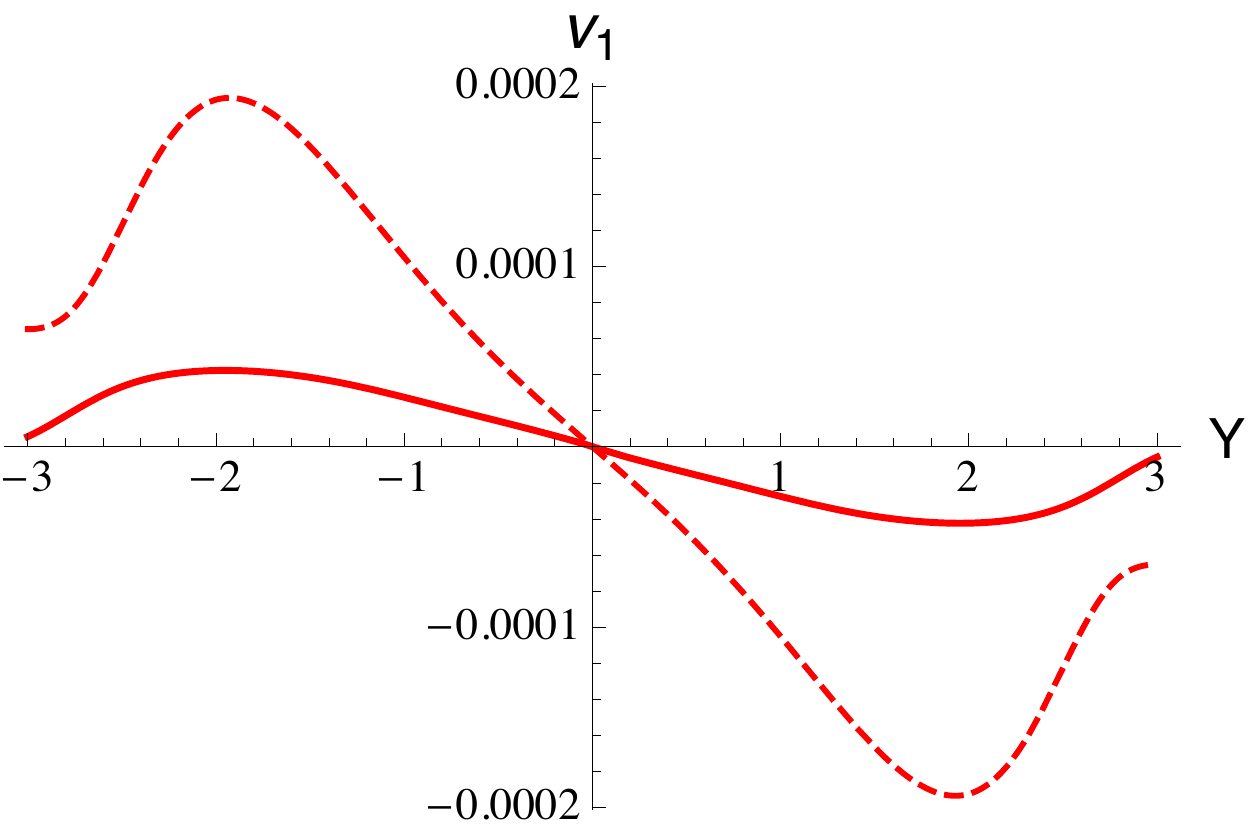}
 \end{center}
\vspace{-.3cm}
 \caption[]{Comparison of Hall vs. Faraday effects. We plot $v_1$ at $p_T=1$~GeV 
as a function of $Y$  for positively charged pions with
 values of parameters chosen as in Fig.~\ref{fig5}, appropriate for LHC collisions.
 The dashed  red curve  is obtained if we turn off the Hall
 effect, keeping only the Faraday effect. The solid red curve, which is the same
 as that in Fig.~\ref{fig5}, includes both the Hall
 and Faraday effects. We see that the Hall and Faraday effects have opposite sign,
 as in Fig.~\ref{fig1}.  Here, the Faraday effect is stronger.}
\label{fig55}
\end{figure}
%%%%%%%%%%%%%%%%%%%%%%%%%%%%%%%%%%%%%%%%%%%%%%%%%%%%%%%% 

We see in Fig.~\ref{fig1} that if the current induced by Faraday's law
is greater than that induced by the Hall effect, we expect $v_1>0$ 
for negative pions at $Y>0$ and for positive pions at $Y<0$
and we expect $v_1<0$ for positive pions at $Y>0$ and for
negative pions at $Y<0$.
Comparing to Fig.~\ref{fig5}, we observe that this is
indeed the pattern for pions with $p_T=1$~GeV, meaning
that in the competition between
the Faraday and Hall effects, the effect of Faraday on pions with $p_T=1$~GeV 
is greater than the effect of Hall.  However,
the effects of Hall and Faraday on pions with smaller $p_T$ and small $Y$
are comparable in magnitude, for example with 
the Hall effect just larger for $p_T=0.25$ and $|Y|<1.2$, resulting in a reversal
in the sign of $v_1$ in this kinematic range.

We can check that
the Faraday and Hall effects make contributions with opposite sign
to the directed flow $v_1$, as illustrated schematically in Fig.~\ref{fig1}. 
In order to calculate the contribution to $v_1$ that is caused by the 
magnetic field only via Faraday's law we
proceed as follows.  We solve for the electric and magnetic
fields in  the center-of-mass frame, as always.  The electric field $E_x$ is that due to
Faraday's law: it is present because $B_y$ is decreasing with time.  So, we compute
a drift velocity $\vec{v}$ (or -$\vec{v}$)  for positively (or negatively) charged particles 
by solving $q \vec E = \mu m \vec{v}$
in the center-of-mass frame.  At each point in space time we then add this $\vec{v}$ (and $-\vec{v}$) 
to the 
charge-independent flow velocity $\vec{u}$ using special relativistic addition of velocities,
and form a four-velocity from the sum.
In this way we obtain $V^{+\mu}$ (and $V^{-\mu}$) that include the velocity from Gubser's flow
as well as the additional velocity for positively (and negatively) charged particles that is induced
by Faraday's law. But, we have left out the Hall effect. We can then compute $v_1$. 
In Fig.~\ref{fig55} we show 
$v_1$ for pions with $p_T=1$~GeV in our calculation with parameters appropriate for
LHC collisions.
The solid curve is the full result, including both the Hall and Faraday effects.
The dashed curve shows the $v_1$ due only to Faraday, with the Hall effect turned
off.  We see that the full result arises from a partial cancellation between the Hall
and Faraday effects, which act in opposite directions as
in Fig.~\ref{fig1}.  For pions with $p_T=1$~GeV, the Faraday effect makes
the larger contribution to $v_1$.
We see, though, that the contribution to $v_1$ due to the Hall current 
is comparable to that arising solely from the Faraday effect.  It would therefore
be interesting to attempt a full-fledged magnetohydrodynamic study in which the
back-reaction of this current on the magnetic field is taken into consideration.
We leave this to future work.

 %%%%%%%%%%%%%%%%%%%%%%%%%%%%%%%%%%%%%%%%%%%%%%%
\begin{figure}[t]
 \begin{center}
\includegraphics[scale=.68]{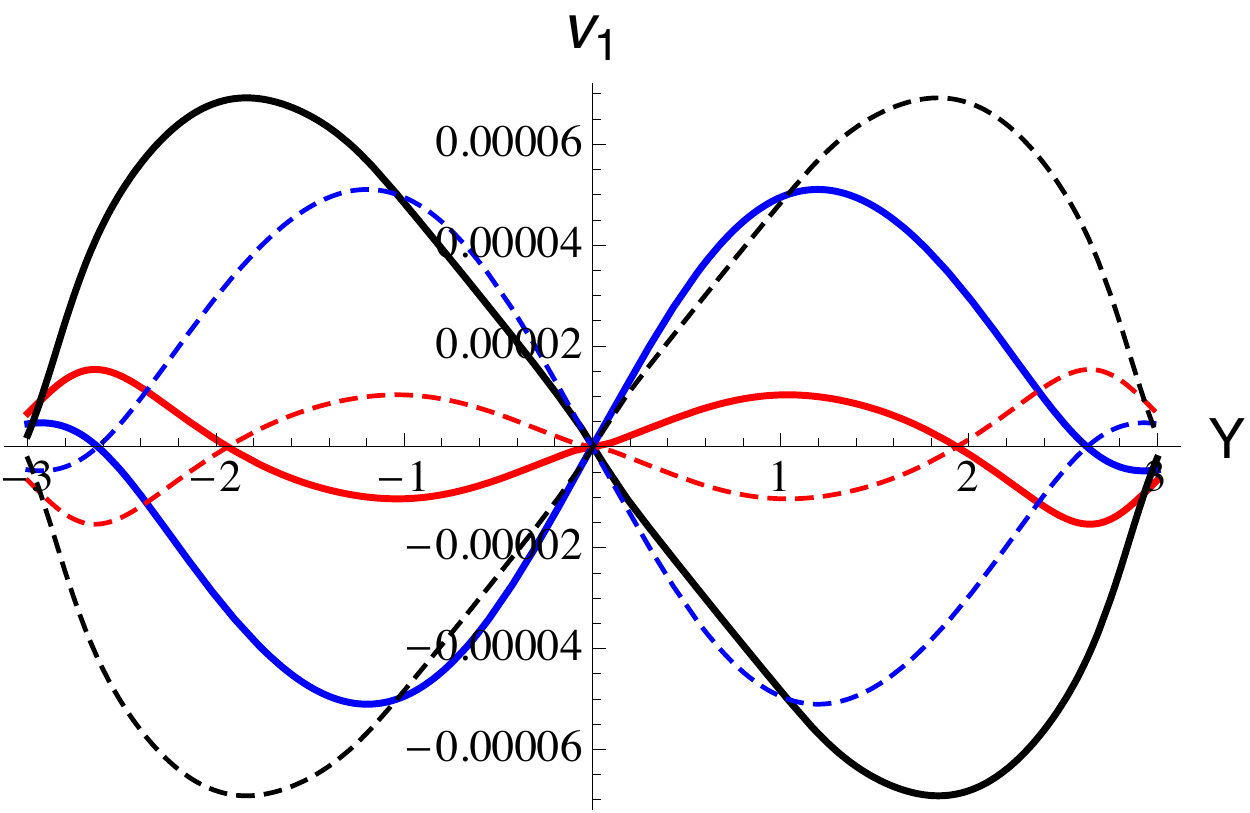}
 \end{center}
\vspace{-.3cm}
 \caption[]{$v_1$ for protons (solid curves) and antiprotons (dashed curves) 
 in our calculation with the same parameters as in Fig.~\ref{fig5}, namely parameters
 chosen 
with 20-30\% centrality heavy ion collisions
at the LHC in mind. We plot $v_1$
as a function of momentum space rapidity $Y$ at $p_T = 0.5$ (blue), 1 (red) and 2 GeV (black). }
\label{fig6}
\end{figure}
%%%%%%%%%%%%%%%%%%%%%%%%%%%%%%%%%%%%%%%%%%%%%%%%%%%%%%%%%%%

Next, we repeat the same calculation as in Fig.~\ref{fig5}, this time 
for the protons and antiprotons.  In Fig.~\ref{fig6} we plot $v_1$ for (anti)protons as  
a function of momentum-space rapidity $Y$ at transverse momenta $p_T = 0.25$, 0.5, and 1 GeV.  
We observe that in the range of parameters $p_T$ and $Y$ that we are interested in $v_1$ for protons turn out to be in they opposite direction to the $v_1$ for pions.  So, when it comes
to their influence on the directed flow of protons in collisions
at LHC energies, the Hall effect is stronger than the Faraday
effect.  
How is it possible for the Faraday effect to be stronger for pions
while the Hall effect is stronger for protons? First, 
in some regions of spacetime the electric current
induced by the Faraday effect 
is greater
than the current induced by the Hall effect whereas in other
regions of spacetime the Hall current is greater.  And, second, 
because $m_T$ is so
much larger for protons than for pions when one computes $v_1$
the integral (\ref{I1}) over the freezeout surface weights the contribution
from different regions of the freezeout surface substantially differently
for protons than for pions.
Putting these together, it turns out that the Hall contribution to $v_1$ for
protons is larger than that from the Faraday effect, whereas it is smaller for the pions.

Interestingly,  the magnitude of $v_1$ is less for protons with $p_T=1$~GeV
than it is at lower $p_T$, meaning that the $p_T$-dependence of $v_1$
for protons in Fig.~\ref{fig6} is opposite that for pions in Fig.~\ref{fig5}.
These observations indicate that for both pions and protons the
magnitude of the Faraday contribution to $v_1$ increases with increasing $p_T$
faster than the magnitude of the Hall contribution.

 %%%%%%%%%%%%%%%%%%%%%%%%%%%%%%%%%%%%%%%%%%%%%%%
\begin{figure}[t]
 \begin{center}
\includegraphics[scale=.7]{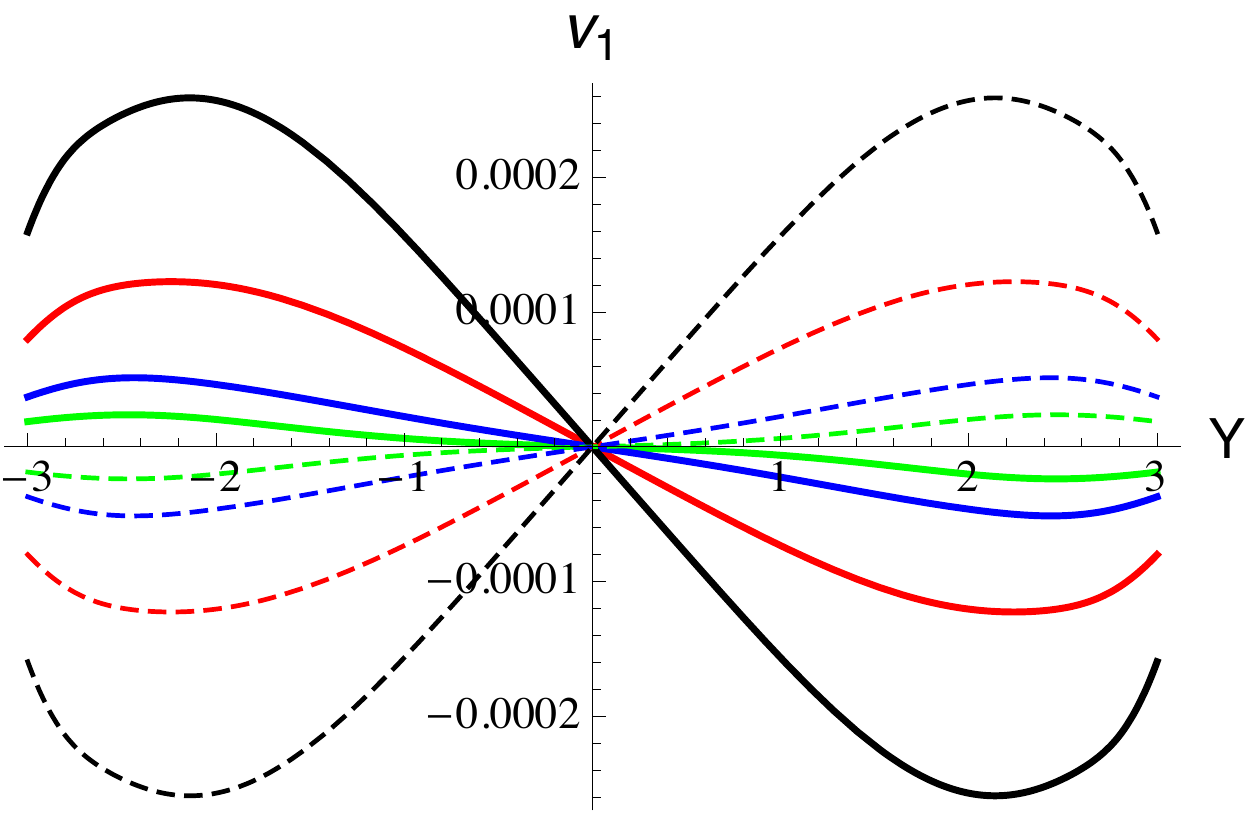}
 \end{center}
\vspace{-.3cm}
 \caption[]{$v_1$ for positively (solid curves) and negatively (dashed curves) charged pions 
 with parameters chosen as for a 20-30\% centrality heavy ion collision at RHIC. We plot $v_1$ 
 as a function of momentum space rapidity $Y$ at $p_T = 0.25$ (green), 0.5 (blue) 1 (red) and 2 GeV (black). Antiprotons are not displayed in this figure for visual clarity.
}
\label{fig7}
\end{figure}
%%%%%%%%%%%%%%%%%%%%%%%%%%%%%%%%%%%%%%%%%%%%%%%%%%%%%%%%%%%

 %%%%%%%%%%%%%%%%%%%%%%%%%%%%%%%%%%%%%%%%%%%%%%%
\begin{figure}[t]
 \begin{center}
\includegraphics[scale=.7]{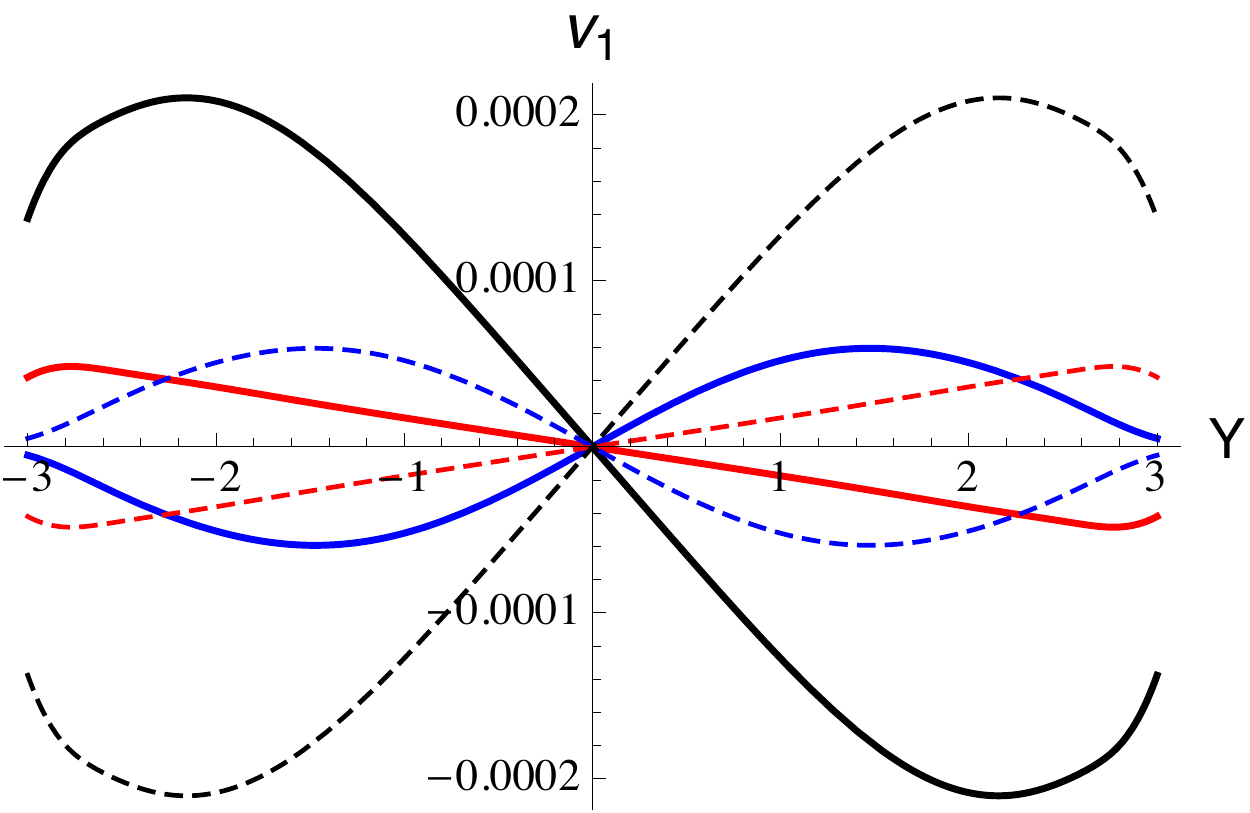}
 \end{center}
\vspace{-.3cm}
 \caption[]{$v_1$ for protons 
 %(solid curves) and
 %antiprotons (negative curves) 
 with parameters chosen as in Fig.~\ref{fig7}, so as to yield estimates for RHIC.
 We plot $v_1$ as a function of $Y$ at $p_T = 0.5$ (blue), 1 (red) and 2 GeV (black).
Anti-protons are not displayed in this figure for visual clarity.}
\label{fig8}
\end{figure}
%%%%%%%%%%%%%%%%%%%%%%%%%%%%%%%%%%%%%%%%%%%%%%%%%%%%%%%%%%%

Finally we present  our estimates for heavy ion collisions at RHIC with $\sqrt{s}=200$~AGeV 
and 20-30\% centrality. 
That is, now we
choose an initial magnetic field
 created by spectators with beam rapidity $Y_0=5.4$,
 we set the parameters specifying Gubser's hydrodynamic solution to
 $\hat T_0=7.5$ and $q^{-1}=5.3$~fm,  we choose the electric conductivity $\sigma$,
 the drag parameter $\mu m$ in (\ref{Lorentz}) 
 and the freezeout temperature  $T_f$  as before.
 One change that we made is that in our calculations with these choices
 of parameters we left out the contribution of the participant protons to
 the magnetic and electric fields from the beginning, computing
 only the effects due to the spectators.
 We made this simplifying choice after having checked that, in our
 previous calculations with parameters appropriate for LHC collisions,
 leaving out the participants makes only a less than 10\% difference to
 the calculated $v_1$'s, in most regions of momentum space much less.
 %
% 
%The various parameters in the analysis above has to be changed. One finds that for a good fit to RHIC data for pions and protons (see e.g. \cite{Adler:2003cb}) one needs to alter the parameters of the Gubser's hydrodynamic solution as  $\hat{T}_0=7.5$ and $q=1/5.3$ fm$^{-1}$ for RHIC. Also the rapidity of the spectators in section \ref{EMT} should now be fixed as $Y_0=5.3$. 
%
In Fig.~\ref{fig7}
we plot
$v_1$ for positively and negatively 
charged pions as a function of $Y$ at $p_T=0.25$, 0.5, 1 and 2 GeV. 
We observe that Faraday effect is dominant for pions at RHIC even for $p_T$ as low as 0.25 GeV.
And, in Fig.~\ref{fig8} 
we present $v_1$ of protons in our calculation with parameters chosen
to mock up a RHIC collision with 20-30\% centrality for protons with
$p_T=0.5$, 1 and 2 GeV.   As in Fig.~\ref{fig6}, we see that the magnitude of
the Faraday contribution to $v_1$ increases with increasing $p_T$.  In Fig.~\ref{fig8} we 
see that the sign of $v_1$ flips as $p_T$ increases, as the Faraday contribution goes from
being smaller than to larger than the Hall contribution.

\section{Observables, and a look ahead}
\label{ObservablesSection}

Our estimates of the magnitude of the charge-dependent directed flow of pions and (anti)protons
in heavy ion collisions at the LHC and RHIC, and their dependence on $Y$ and $p_T$,
can be found in Figs.~\ref{fig5}, \ref{fig6}, \ref{fig7} and \ref{fig8}.  If we focus on $Y\sim 1$
and $p_T\sim 1$~GeV, we see that the magnitude of the contribution to $v_1$ due to
the magnetic field is between $10^{-5}$  and $10^{-4}$, with the effect being about twice
as large at in heavy ion collisions at  top RHIC energies than in those at the LHC
and about twice as large for pions than for (anti)protons.
So, the effect is small.  What makes it distinctive is that it is opposite in sign for positively
and negatively charged particles of the same mass, and that for any species it is odd
in rapidity.  
%One can see from Fig. \ref{fig7} that the expected difference between the directed flows of positive and negative hadrons originating from the effects of magnetic field is about $v_1^+ - v_1^- \sim 10^{-4}$. While this is a small difference,  it is detectable.  
Detecting the effect directly by measuring the directed flow of positively
and negatively charged particles, which we shall denote by $v_1^+$ and $v_1^-$, is possible in principle
but is likely to be prohibitively difficult in practice for two reasons.
First, event-by-event there can be significant charge-independent contributions to $v_1$
due to event-by-event variation in the ``shape'' (in the transverse plane) 
of the energy deposited by the collision.  This means that a separate measurement of 
$v_1^+$ and $v_1^-$ followed
by subtracting one measured quantity from the other would 
%involve looking for the difference between
%two quantities that are each separately much larger than their difference.  This
%would 
require enormous data sets and very precise control of each
of the two separate measurements.  
Second, the separate
measurement of either $v_1^+$ or $v_1^-$ requires reconstructing the direction
of the magnetic field in each event (i.e.~determining event by event whether $B_y$ is positive or negative) by using
forward detectors to measure
the directions in which the remnants of the colliding ions are deflected.  
%In experiment, the direction of magnetic field can be reconstructed from the directions in which the remnants of the colliding ions are deflected -- this requires the use of forward detectors (below we also propose observables that do not require the knowledge of the direction of magnetic field).   
It would be advantageous to define correlation observables that, 
first of all, involve taking ensemble averages of suitably chosen differences 
rather than just of $v_1^+$ or $v_1^-$ and that, second of all,
do not require
knowledge of the direction of the magnetic field.  The construction of such
observables 
can be guided by symmetry considerations that apply in collisions between like nuclei
 that dictate that  
%{\it in every single event}, 
the contribution to the directed flow that is caused by the electric currents induced
by a magnetic field created in the collision must satisfy
\be\label{v1ch}
v_1^+(Y) =  - v_1^-(Y) = - v_1^+(-Y) = v_1^-(-Y) \
\ee
for either pions or (anti)protons, for any value of $p_T$,
and regardless of the direction of the magnetic field.
Here $Y=0$ means particles produced at $90^\circ$ to the beam direction
in the center-of-mass frame.

%Since there are also charge-independent effects on the directed flow, to isolate the effects of the charged currents it is beneficial  
%to study in experiment the following 

To isolate the charge-dependent directed flow that we are after, namely the effect of
an electric current as in Fig.~\ref{fig1} that must satisfy (\ref{v1ch}) event-by-event,
and to separate it from larger charge-independent effects it is helpful to define
the following 
asymmetries between the directed flows for positive and negative hadrons:
\bea
A_1^{+-}(Y_1, Y_2) &\equiv&   v_1^+(Y_1) - v_1^-(Y_2),\nn\\ 
A_1^{++}(Y_1, Y_2) &\equiv&   v_1^+(Y_1) - v_1^+(Y_2), \nn\\
A_1^{--}(Y_1, Y_2) &\equiv&   v_1^-(Y_1) - v_1^-(Y_2) ,
\eea
and to measure correlations of these asymmetries.
It is easy to see from (\ref{v1ch}) that for the effects induced by a magnetic field
\bea\label{corr}
A_1^{+-} (Y, Y) &=& 2 v_1^+(Y) = - A_1^{+-} (-Y, -Y)\nn\\ 
{} &=& A_1^{++} (Y, -Y) = A_1^{--} (-Y, Y) \, ,
\eea
and so on. 
Even if the direction of the magnetic field is not reconstructed, one can still study the correlation functions defined by 
\be
C_1^{i, j}(Y_1, Y_2) \equiv \langle A_1^i (Y_1, Y_2) A_1^j (Y_1, Y_2) \rangle .
\ee
These correlation functions are quadratic in the directed flow, and so are not sensitive to the direction of $\vec B$
and the sign of $v_1$ in a given event. 
However, they still carry the requisite information about dynamical charge-dependent correlations
induced by the magnetic field. Analogous correlations functions have been measured with 
high precision~\cite{Abelev:2009ac,Abelev:2009ad}.
Using the relations (\ref{v1ch}) and (\ref{corr}), one can easily construct the desired correlators, 
and can then predict their signs and magnitudes using our results from Section~\ref{ElectricCurrentSection}. 
Let us list four examples. First, consider
\bea
C_1^{+-,+-}(Y, Y) &\equiv& \langle A_1^{+-}(Y, Y) A_1^{+-}(Y, Y) \rangle \nn\\
&=& 4 \langle v_1^+(Y) v_1^+(Y) \rangle ,
\label{FirstCorrelator}
\ee
where we have used (\ref{v1ch}) in the second equality. Charge-independent
contributions to $v_1$ that do not satisfy (\ref{v1ch}) will cancel in (\ref{FirstCorrelator}).
Second, in addition to measuring $C_1^{+-,+-}(Y,Y)$ with the goal of extracting
$\langle v_1^+(Y) v_1^+(Y) \rangle$ it is very important at the same time to measure
\bea
C_1^{+-,+-}(Y, -Y) &\equiv& \langle A_1^{+-}(Y, -Y) A_1^{+-}(Y, -Y) \rangle\nn\\
&=& 0 ,
\label{NextCorrelator}
\ee
since according to (\ref{v1ch}) this correlator should vanish, as indicated by the last equality.  One can of course
also measure
\be
\langle (v_1^+(Y) + v_1^-(-Y))^2\rangle = 4 \langle v_1^+(Y) v_1^+(Y) \rangle ,
\label{NextNextCorrelator}
\ee
where we have used (\ref{v1ch}) in the equality.
Fourth, consider
\bea
C_1^{++,--}(Y, - Y) &\equiv& \langle A_1^{++}(Y, - Y) A_1^{--}(Y, - Y) \rangle \nn\\
&=& 2 \langle v_1^+(Y)v_1^-(Y)   -   v_1^+(Y) v_1^-(-Y)  \rangle \nn\\
&=& - 4 \langle v_1^+(Y) v_1^+(Y) \rangle ,
\label{SecondCorrelator}
\eea
where we have used (\ref{v1ch}) in the last equality.  So, to give an example
of a possible analysis strategy, imagine measuring the four correlators (\ref{FirstCorrelator}),
(\ref{NextCorrelator}), (\ref{NextNextCorrelator}) and (\ref{SecondCorrelator}) in heavy
ion collisions at RHIC or the LHC, for pions or for (anti)protons or for that matter 
for charged hadrons.  Contributions  to these correlators arising from the electric
current induced by the Hall and Faraday effects due to the presence of a magnetic
field will vanish in (\ref{NextCorrelator}), will be equal in (\ref{FirstCorrelator}) 
and (\ref{NextNextCorrelator}), and will be equal in magnitude but opposite
in sign in (\ref{SecondCorrelator}).  Measuring correlations that fit this pattern
will allow for the determination of $\langle v_1^+(Y) v_1^+(Y) \rangle$, which could
then be compared to the results of calculations like those we have presented in
Section~\ref{ElectricCurrentSection}.

Finally, it may also be advantageous
to measure the components of (\ref{SecondCorrelator}), namely
$\langle v_1^+(Y) v_1^-(Y)\rangle$ and $\langle v_1^+(Y) v_1^-(-Y)\rangle$, separately.
Measuring each of these correlators and showing that they are both nonzero, are
equal in magnitude, and that the first is negative while the second is positive would also
constitute strong evidence for
 the charge-dependent and rapidity-odd contribution to the directed flow
induced by the magnetic field present during the collision.

The challenge to experimentalists is to measure these correlators, or others that are 
also defined so as to separate the effects that satisfy (\ref{v1ch}) from charge-independent
backgrounds.  If this is possible, one can
imagine that it may be possible to use comparisons between data and 
the nontrivial $p_T$- and $Y$-dependence of
results like those that we have obtained in Figs.~\ref{fig5}, \ref{fig6}, \ref{fig7} and \ref{fig8}
to extract a wealth of information, for example about the strength of the initial magnetic field and about
the magnitude of the electrical conductivity of the plasma.  

Before such goals can be realized, however, there remain many challenges
on the theoretical side.  We have made many simplifying assumptions, justifying
them by virtue of the fact that our goal in this paper is only order-of-magnitude
estimates of the Hall and Faraday effects on the charge-dependent directed flow.
Given that the magnitude of the observable effect turns out to result from a partial
cancellation between the Hall and Faraday effects, and given the interesting and quite nontrivial 
dependence of our results on $Y$ and $p_T$, there is plenty of motivation for a more
sophisticated, less simplified, treatment.  
In our view, the most pressing challenges
are the inclusion of temperature-dependent, and therefore spacetime-dependent,
electrical conductivity $\sigma$ and drag parameter $\mu m$, as well as the {\it ab initio}
calculation of the second of these two quantities.  Treating both these quantities as
temperature-dependent, rather than as constants, will require an analysis in which
the solution of Maxwell's equations is done numerically, rather than analytically
as in Section~\ref{EMT}.  Once this threshold has been crossed, there will
be no motivation to use Gubser's analytic solution to the hydrodynamic equations.  At this
point it will be best to use a state-of-the-art (3+1)-dimensional numerical 
relativistic viscous hydrodynamics
calculation.
Even further in the future it may become relevant to consider the back reaction of the 
effects induced by the magnetic field on the hydrodynamics itself.  However, given the 
smallness of the effects that we have found, attempting this even more challenging extension of
our analysis does not seem to be pressing.

A natural direction
for further investigation is lower energy heavy ion collisions, as
in the RHIC Beam Energy Scan program. Heavy ion
collisions with $\sqrt{s}$ as low as 7.7 AGeV have been studied
in the first, exploratory, phase of this program.  The STAR collaboration
has measured the directed flow $v_1$ for positively and negatively
charged pions and for protons and antiprotons in these
lower energy collisions~\cite{Pandit:2011np}. These preliminary
data show hints of the effects of magnetic fields that we
have described, for example 
with $v_1$ for positively charged pions less than (greater than)
$v_1$ for negatively charged pions
with $Y>0$ ($Y<0$),
as when the Faraday effect dominates
over the Hall effect,  in collisions with $\sqrt{s}=7.7$ and 11.5 AGeV.
This
motivates the measurement of the directed
flow correlations that we have proposed.
High statistics data sets at these low collision energies are anticipated
in a few years, after the implementation of a RHIC
upgrade involving adding electron cooling for
lower energy heavy ion beams.

Since we have found that the observable effects of 
magnetic fields on the charge-dependent directed flow are greater 
at top RHIC energies than at LHC energies, 
it is natural to expect that
the effects will be greater still in lower energy collisions at RHIC.
At these lower energies, however, the calculation of these effects
is much more challenging for several reasons.
The matter produced in the collision spends
less time in the QGP phase meaning that it spends a larger fraction
of its time in the vicinity of the crossover or transition between QGP
and hadron gas, and in the hadron gas phase.  This makes the use
of a constant $\sigma$ and the use
of a solution to conformal hydrodynamics, like Gubser's, less viable
even as qualitative guides.  We look forward to estimating
the magnitude of the directed flow correlators that we have introduced 
in this paper in 
lower energy collisions in the future, once a treatment with $\sigma$ varying
in space and time and with more realistic hydrodynamics is in hand.
Also, at the lowest energies the assumption that
we made in calculating the magnetic field that the spectators travel
along straight lines will no longer be valid. Finally, the assumption that
all the fragments of the incident nucleons (spectators
and participants) end up at large $|Y|$, well separated from
the smaller values of $|Y|$ where we look for effects of the magnetic
field, must also break down in lower energy collisions with
smaller beam rapidity. For all these reasons, 
further calculations are needed before firm conclusions can be drawn
from the low energy data. And, there are strong motivations
for measuring the directed flow correlators that we have
defined in heavy ion collisions at top RHIC energy and at the LHC, where
our estimates should be more reliable.

%\section{Summary}
%\lab{summary}

%\\ \\
{\bf Acknowledgements.} 
We are grateful to Sergei Voloshin for helpful suggestions.
U.G.~and K.R.~are grateful to the CERN Theory division
for hospitality at the time this research began. 
The work of D.K.~was
supported in part by the U.S. Department of Energy under Contracts No. DE-FG-88ER40388 and
DE-AC02-98CH10886. 
The work of K.R.~was supported by the U.S. Department of Energy
under cooperative research agreement DE-FG0205ER41360. This work is part of the D-ITP consortium, a program of the Netherlands Organisation for Scientific Research (NWO) that is funded by the Dutch Ministry of Education, Culture and Science (OCW). 

\vskip 0.5cm

\vskip 1cm
%\newpage

%\end{narrowtext}

\end{document}